\documentstyle [12pt]{article}

\begin{document}

\begin{titlepage}

\begin{flushright}
IFFC-93-07 \\  September 1993
\end{flushright}

\vspace{\fill}

\begin{center}
    {\large \bf SU(2) QCD IN THE PATH REPRESENTATION: \\[0.15cm]
                GENERAL FORMALISM AND MANDELSTAM      \\[0.15cm]
                INDENTITIES}
\end{center}

\vspace{1.0cm}

\begin{center}
     {\large Rodolfo Gambini and Leonardo Setaro}
\end{center}

\begin{center}
         {\em  Instituto de F\'{\i}sica, Facultad de Ciencias \\
               Trist\'{a}n Narvaja 1674, Montevideo, Uruguay}
\end{center}

\vspace{\fill}

\renewcommand{\thesection}{\Roman{section}}


\begin{abstract}

We introduce a path-dependent hamiltonian representation (the path
representation) for SU(2) with fermions in 3 + 1 dimensions. The
gauge-invariant operators and hamiltonian are realized in a Hilbert
space of open path and loop functionals. We obtain two new types of
Mandelstam identities one that connects open path operators with
loop operators and other involving the end points of the paths. \\

\end{abstract}

\end{titlepage}

\newpage


\section{Introduction}
\label{sec:intro}

\ \\
For many years, considerable effort has been made to understand the
nume\-rical aspects of lattice gauge theories. However, in spite of
many remarkable achievements, these calculations have some inherent
limitations and they do not help to understand many of the
beautiful analytic features of the theory. The Montecarlo methods
are probably not the most economical device to understand the real
dynamics of the system. In fact, the highly symmetric nature of the
action makes the numerical treatments of the Euclidean partition
function somewhat inefficient. Furthermore, computational
simulations including dynamical fermions are very expensive in
computer time due to the Grassmannian character of the fermionic
variables. \\

A first natural step towards the identification of the real degrees
of freedom of the problem is to use the hamiltonian approach. The
residual time independent gauge symmetry can be eliminated with the
introduction of gauge invariant objects. Because of their simple
behavior under gauge transformations, holonomies or phase factors
may be used as the basic variables, without ever resorting to the
use of gauge potentials or field strengths. The hamiltonian
approach to pure lattice gauge theories in terms of a basis of
gauge invariant loop states is known as the loop representation
\cite{Gam86}. This approach has recently deserved much attention
due to the work of Rovelli and Smolin \cite{Rov88,Rov90} on the
$SL(2,C)$ gauge theory formulation of quantum general
relati\-vi\-ty. That is why several alternative proposals to
formulate the lattice gauge theory in a gauge invariant manner have
recently appeared \cite{Bru91,Lol92a,Lol92b,Lol93a,Lol93b}. \\

The use of the loop representation as a tool to extract information
from a hamil\-tonian eigenvalue equation has been considered by
many authors \cite{Bru91,DiB89,Fur87,Gam89,Aro} mainly in the case
of pure abelian and non abelian $SU(2)$ and $SU(3)$ gauge theories.
The action of the hamiltonian operator typically results in
geometric deformations or rearrangements (fusion, fission,
rerouting, etc) of the loops. In order to introduce matter into the
formalism, loops need to be ge\-ne\-ralized to open paths with
charged fermions at the end points. The path representation has
been recently applied to the study of quantum electrodynamics with
light fermions \cite{For91} leading to very interesting results
about the phase structure of the hamiltonian theory
\cite{For92a,For92b}. \\

In this paper, we extend the path representation to the case of
lattice QCD with staggered fermions. As it was already noticed
\cite{For91} in the case of QED the open path representation
presents clear advantages when fermions are present. In fact, the
main features of the loop representation for pure gauge theories
are preserved. The Grassmannian character of the fermionic
variables is absorbed into the geometry of the gauge invariant path
func\-tio\-nals and the introduction of fermions only implies a
moderate increment in the dimensionality of the Hilbert space. The
basis of the Hilbert space in the path representation is naturally
associated with the physical excitations (gluons, mesons and
barions) of the theory. The loop representation generally involve
an overcomplete set of basis elements labeled by loops. In order to
have a representation equivalent to the original connection
representation, one has to impose appropiate conditions on the set
of loop dependent functions. The conditions usually called the
Mandelstam constraints are a set of group dependent relations among
the loop variables \cite{Gil81,Gam86,Lol92a,Lol93a}. In the case of
gauge theories in presence of matter we have two new types of
Mandelstam constraints, one that relates open path dependent
functions with loop dependent functions and other that involves the
end points of the paths. These constraints are related with the
reconstruction properties of the spinorial variables and play a
fundamental role at the physical level. In fact we shall see that
in this language the QCD hamiltonian only involves mesonic
operators and the barionic states only arise via the new Mandelstam
constraints. Although the formulation is ge\-ne\-ral in scope we
have restricted ourselves in this paper to the $SU(2)$ QCD. \\

The paper is organized as follows. In section~\ref{sec:alg} we
introduce the path representation for $SU(2)$ with fermions in
\mbox{3 + 1} dimensions and we study the realization of the
operators in a Hilbert space. In section~\ref{sec:Man-Id} we
introduce the new Mandelstam identities for open paths and we show
how this constraints may be used to reduce the redundancy of the
basis of paths. In section~\ref{sec:Ham-Dyn} we examine the
symmetries and action of the hamiltonian in the path representation
and we sketch how the path dependent formalism allows to accomplish
explicit calculations. Conclusions and some final remarks are given
in section~\ref{sec:Con}.


\section{Algebra of the Gauge Invariant Path Functionals}
\label{sec:alg}

\indent The formulation of a quantum gauge theory in the Hilbert
space of kets $\mid \Psi , A_{\mu} \rangle$ require to work with a
multiplicity of equivalent states under gauge transformations. This
redundancy in the states may be eliminated if we work in a
path-dependent hamiltonian representation, the $P$
representation, which offers a direct gauge-invariant
characterization of the physical states by a basis of the kets
$\mid {\bf P} \rangle$, where {\bf P} labels a set of open paths
and loops, associate to lines of flux and pure excitations. The
gauge invariant objects that one can construct from the phase
factor $U(P)$ and the fermionic fields are given by

\begin{eqnarray}
W(L) & = & Tr \left[ U(L) \right] \nonumber \\
\Omega_{ij}^{fg} \left( P_{x}^{\ y} \right) & = &
  {\Psi^{\dagger}}_{iA}^{f} (x) U_{AB}(P_{x}^{\ y})
                                         \Psi_{jB}^{g}(y) \\
\Xi_{ij}^{fg} \left( P_{x}^{\ y} \right) & = &
  \Psi_{iA}^{f}(x) \varepsilon_{AB} U_{BC}(P_{x}^{\ y})
                                             \Psi_{jC}^{g}(y)
                                            \nonumber
\end{eqnarray}

\noindent where $L$ is a loop and the fermionic fields are located
at the ends of the open path P. The indices $f$ and $g$ label the
flavours of the quarks, $i$ and $j$ are spinorial indices and $A,B
= 1,2$ are the internal color indices for the representation of
SU(2). The difficulty in changing to the $P$ representation is
that the scalar product between the former configuration vectors
and {\bf P} vectors is not defined. In fact the gauge invariant
object $\Omega$ is not a pure object of the configuration space
because it includes the canonical conjugate moment of $\Psi$. The
solution is to consider the Hilbert space of kets $\mid
\Psi^{\dagger}_{u} , \Psi_{d} , A_{\mu} \rangle$, where $u$
corresponds to the upper part of the Dirac spinor and $d$ to the
lower part, and to define the scalar inner product with a {\bf P}
vector as

\begin{eqnarray}
\Phi_{o} \left( P_{x;i,j}^{\ y;f,g} \right) & \equiv &
          \langle P_{x;i,j}^{\ y;f,g}
           \mid \Psi^{\dagger}_{u} , \Psi_{d} , A_{\mu} \rangle
                                                     \nonumber \\
& = & {\Psi^{\dagger}_{u}}_{iA}^{f}(x) U_{AB}(P_{x}^{\ y})
                                    {\Psi_{d}}_{jB}^{g}(y)
                                                     \nonumber \\
& = & {\Psi^{\dagger}_{u}}_{iA}^{f}(x)
   {\left\{ exp \left[ -i \int_{x}^{y} A_{i}(y,t) \, dy_{i}
                \right]_{ord} \right\}}_{AB} {\Psi_{d}}_{jB}^{g}(y)
\end{eqnarray}

\noindent We have also the following gauge invariant path operators

\begin{eqnarray}
\Phi_{1} \left( P_{x;i,j}^{\ y;f,g} \right) & = &
    {\Psi^{\dagger}_{u}}_{iA}^{f}(x) U_{AB}(P_{x}^{\ y})
                                  {\Psi_{u}}_{jB}^{g}(y)
                                                   \nonumber \\
\Phi_{2} \left( P_{x;i,j}^{\ y;f,g} \right) & = &
    {\Psi^{\dagger}_{d}}_{iA}^{f}(x) U_{AB}(P_{x}^{\ y})
                                  {\Psi_{d}}_{jB}^{g}(y)     \\
\Phi_{3} \left( P_{x;i,j}^{\ y;f,g} \right) & = &
    {\Psi^{\dagger}_{d}}_{iA}^{f}(x) U_{AB}(P_{x}^{\ y})
                                  {\Psi_{u}}_{jB}^{g}(y)
                                                   \nonumber
\end{eqnarray}

\begin{equation}
W(L) = Tr \left[ U(L) \right]
     = Tr \left\{ exp \left[ -ig \oint_{L} A_{i}(y,t) dy_{i}
                           \right]_{ord} \right\}  \label{Fcont}
\end{equation}

In the lattice we consider the configuration basis $\{ \mid
\chi^{\dagger} (x_e), \chi (y_{o}) , U(\ell) \rangle \}$, where the
$\chi$'s are single-component lattice fields, or staggered fermions
\cite{Ban77}, and $x_{e}$ and $y_{o}$ labels even and odd sites in
the net. The $\chi$ field carries a color index and obeys the
canonical anticommutation relations

\begin{equation}
\left[\chi^{\dagger} (\vec{r}) ,\chi (\vec{r}\,')\right]_{+}
        = \delta_{\vec{r},\vec{r}'}              \hspace{1.0cm}
\left[\chi (\vec{r}) ,\chi (\vec{r}\,')\right]_{+} = 0
                                                  \label{cChCh}
\end{equation}

We define the path operator $\tilde U (P_{x}^{\ y})$ as the product
of link operators $U(\ell)$, with the weight factors $\eta(\ell)$,
along of the path $P_{x}^{\ y}$

\begin{equation}
\tilde U (P_{x}^{\ y}) = \prod_{\ell \in P_{x}^{\ y}} U(\ell)
                                  \eta (\ell)
\end{equation}

\noindent the $\eta$'s are

\begin{eqnarray}
 \eta(\widehat{x}_{1}) = (-1)^{x_{3}} \hspace{0.5cm}
 \eta(\widehat{x}_{2}) = (-1)^{x_{1}} \hspace{0.5cm}
 \eta(\widehat{x}_{3}) = (-1)^{x_{2}} \hspace{0.5cm}
 \eta(- \widehat{n}) = \eta(\widehat{n})        \nonumber \\
\end{eqnarray}

\noindent where $(x_{1}, x_{2}, x_{3})$ are the origin coordinates
of the link $\ell$. \\

The factors $\eta(\ell)$ arise when we work with staggered fermions
\cite{For91}, they have the information of the Dirac matrices in
the lattice, and their introduction in the $\tilde U (P_{x}^{\ y})$
operator allows to simplify the expression of the hamiltonian and
its symmetry properties. \\

Negative links are not independent since one demands

\begin{equation}
U(\overline{\ell}) = U^{\dagger}(\ell)       \label{Udaga}
\end{equation}

The inner products between configuration space vectors and path
dependent vectors are given by

\begin{eqnarray}
\langle \chi^{\dagger}(x_e), \chi(y_{o}) , U(\ell)
                             \mid L \rangle
  & = & Tr \left[ U(L) \right]                     \nonumber \\
\langle \chi^{\dagger}(x_e), \chi(y_{o}) , U(\ell)
                             \mid P_{x_{e}}^{y_{o}} \rangle
  & = & \chi_{_A}^{\dagger}(x_{e}) \,
             \tilde {U}_{_{AB}} (P_{x}^{\ y})
                             \, \chi_{_B}(y_{o})
                                                              \\
\langle \chi^{\dagger}(x_e), \chi(y_{o}) , U(\ell)
                                  \mid P_{x_{o}}^{y_{o}} \rangle
  & = & \chi_{_A}(x_{o}) \, \varepsilon_{_{AB}} \,
     \tilde {U}_{_{BC}} (P_{x}^{\ y}) \, \chi_{_C}(y_{o})
     \nonumber \\
\langle \chi^{\dagger} (x_e), \chi (y_{o}) , U(\ell)
                                  \mid P_{x_{e}}^{y_{e}} \rangle
  & = & \chi_{_A}^{\dagger} (x_{e}) \,
    \tilde {U}_{_{AB}} (P_{x}^{\ y})
         \, \varepsilon_{_{BC}} \, \chi_{_C}^{\dagger}(y_{e})
                                                   \nonumber
\end{eqnarray}

\noindent where $L$ is a loop and the paths $P_{x_{e}}^{\ y_{o}}$
connect one antiquark with one quark, $P_{x_{o}}^{\ y_{o}}$ two
quarks, and $P_{x_{e}}^{\ y_{e}}$ two antiquarks. We choose the
even sites, $x_{e}$, as the beginning and odd site, $y_{o}$, as the
end of the paths between one antiquark and one quark, later we
shall explain the reason for this choice. Then in this
representation we introduce the following set of path dependent
operators:

\begin{equation}
\tilde{W}(L) =
      Tr\left[ \prod_{\ell \in L} U(\ell) \eta (\ell) \right]
\end{equation}

\begin{eqnarray}
\phi_{0}(P_{x}^{\ y}) & = & \chi_{_A}^{\dagger}(x_{e}) \,
             \tilde{U}_{_{AB}}(P_{x}^{\ y}) \, \chi_{_B}(y_{o})
                                                 \nonumber \\
\phi_{1}(P_{x}^{\ y}) & = & \chi_{_A}^{\dagger}(x_{e}) \,
            \tilde{U}_{_{AB}}(P_{x}^{\ y}) \, \chi_{_B}(y_{e}) \\
\phi_{2}(P_{x}^{\ y}) & = & \chi_{_A}^{\dagger}(x_{o}) \,
             \tilde{U}_{_{AB}}(P_{x}^{\ y}) \, \chi_{_B}(y_{o})
                                                 \nonumber \\
\phi_{3}(P_{x}^{\ y}) & = & \chi_{_A}^{\dagger}(x_{o}) \,
             \tilde{U}_{_{AB}}(P_{x}^{\ y}) \, \chi_{_B}(y_{e})
                                                 \nonumber
\end{eqnarray}

\noindent We can write the four operators $\phi_{i}$ like

\begin{equation}
\phi(P_{x}^{\ y}) = \chi_{_A}^{\dagger}(x) \,
                 \tilde{U}_{_{AB}}(P_{x}^{\ y}) \, \chi_{_B}(y)
\end{equation}

\noindent where the sites may be even or odd. The operator
$\tilde{W}(L)$ is associated with the Wilson operator through

\begin{equation}
\tilde{W} (L) = Tr\left[\prod_{\ell\in L}U(\ell)\eta(\ell)\right] =
            \left[\prod_{\ell\in L}\eta(\ell)\right] \ W(L)
\end{equation}

\noindent and for the single plaquette we have

\begin{equation}
\tilde{W}(\Box) = - W(\Box)               \label{WW}
\end{equation}

We shall have another set of four gauge-invariant operators

\begin{equation}
\Gamma(P_{x}^{\ y}) =
\chi_{_A}(x) \, \varepsilon_{_{AB}} \,
        \tilde{U}_{_{BC}}(P_{x}^{\ y}) \, \chi_{_C}(y)
\end{equation}

\noindent The operator $\tilde{W}$ will be associated to gluon
excitations. The operator $\phi_{0}$ represents the quark-antiquark
interaction, a meson, and

\begin{eqnarray}
\Gamma_{2} (P_{x}^{\ y}) & = & \chi_{_A}(x_{o}) \,
\varepsilon_{_{AB}}
              \, \tilde{U}_{_{BC}}(P_{x}^{\ y}) \,
              \chi_{_C}(y_{o}) \\
\Gamma_{1}^{\dagger}(P_{x}^{\ y}) & = &
     - \chi_{_A}^{\dagger}(y_{e}) \, \tilde{U}_{_{AB}}(P_{y}^{\ x})
                \, \varepsilon_{_{BC}} \, \chi_{_C}^{\dagger}(x_{e})
\end{eqnarray}

\noindent are respectively associated to quark-quark and
antiquark-antiquark interactions, which are the analogous of
barions and antibarions in SU(2) \cite{Bow}. We shall call
$\phi_{0}$ a {\em mesonic operator} and $\Gamma_{2}$ and
$\Gamma_{1}^{\dagger}$ {\em barionic operators}. \\

We have also the electric operator

\begin{equation}
E_{\ell} = E_{\ell}^{\ A} X^{A}
\end{equation}

\noindent which satisfies the commutation relations

\begin{eqnarray}
\left[ E_{\ell}^{\ A}, \ E_{\ell '}^{\ B} \right]
    & = & i \, \varepsilon^{ABC} \, \delta_{\ell \ell '}
    \, E_{\ell}^{\ C}
                                                 \label{cEE} \\
\left[ E_{\ell}^{\ A}, \ \tilde{U}_{\ell '} \right]
    & = & -\delta_{\ell \ell '} \, X^{A} \, \tilde{U}_{\ell}
                                                  \label{cEU}
\end{eqnarray}

\noindent where $\varepsilon^{^ABC}$ are the structure constants
and $X^{A}$ the generators

\begin{equation}
\left[ X^{A}, \ X^{B} \right] = i \, \varepsilon^{^ABC} \, X^{C}
\end{equation}

\noindent which are normalized by

\begin{equation}
Tr(X^{A} \, X^{B}) = \delta^{AB}
\end{equation}

The algebra of this gauge invariant representation is obtained from
the commutators (\ref{cEE}) and (\ref{cEU}), the anticommutators
(\ref{cChCh}), and from the following equal time commutators of the
bosonic and fermionic operators

\begin{equation}
\left[ E_{\ell}^{\ A}, \ \chi (x) \right] =
\left[ E_{\ell}^{\ A}, \ \chi^{\dagger}(x) \right] =
\left[ \tilde{U}_{\ell}, \ \chi (x) \right] =
\left[ \tilde{U}_{\ell}, \ \chi^{\dagger}(x) \right] = 0
\end{equation}

\noindent Then, the algebra of the gauge invariant operators is
given by
\begin{eqnarray}
\left[ \phi (P_{x}^{\ y}) ,
                      \phi ({P'}_{u}^{\ v}) \right]
  & = & \delta_{yu} \phi (P_{x}^{\ y}  {P'}_{u}^{\ v})
             - \delta_{vx} \phi ({P'}_{u}^{\ v} P_{x}^{\ y})
                                               \label{cphiphi} \\
\left[ \Gamma (P_{x}^{\ y}) ,
                      \Gamma ({P'}_{u}^{\ v}) \right] & = & 0 \\
\left[ \Gamma (P_{x}^{\ y}) ,
          \Gamma^{\dagger} ({P'}_{u}^{\ v}) \right]
  & = & \delta_{yv} \delta_{xu}
         \tilde{W} \left( {P'}_{u}^{\ v} P_{y}^{\ x} \right) +
                        \delta_{yu} \delta_{xv}
         \tilde{W} \left( {P'}_{v}^{\ u} P_{y}^{\ x} \right)
                                                   \nonumber \\
  &   & - \delta_{yv}
          \phi \left( {P'}_{u}^{\ v} P_{y}^{\ x} \right) -
                             \delta_{yu}
          \phi \left( {P'}_{v}^{\ u} P_{y}^{\ x} \right)
                                                   \nonumber \\
  &   & - \delta_{xv}
          \phi \left(P_{y}^{\ x} {P'}_{v}^{\ u} \right) -
         \delta_{xu}
          \phi \left( {P'}_{v}^{\ u} P_{x}^{\ y} \right)     \\
\left[ \phi (P_{x}^{\ y}) , \Gamma ( P_{u}^{\ v}) \right]
  & = & - \delta_{xv} \Gamma (P_{u}^{\ v} P_{x}^{\ y})
          - \delta_{ux} \Gamma (P_{y}^{\ x} P_{u}^{\ v})     \\
\left[ \phi (P_{x}^{\ y}) ,
                       \Gamma^{\dagger} ( P_{u}^{\ v}) \right]
  & = & \delta_{uy} \Gamma^{\dagger} ( P_{x}^{\ y} P_{u}^{\ v})
       + \delta_{vy} \Gamma^{\dagger} (P_{x}^{\ y} P_{v}^{\ u})
\end{eqnarray}

\begin{equation}
\left[ \phi   (P_{x}^{\ y}), \tilde{W}(L) \right] =
\left[ \Gamma (P_{x}^{\ y}), \tilde{W}(L) \right] =
\left[ \Gamma^{\dagger} (P_{x}^{\ y}), \tilde{W}(L) \right] = 0
                                               \label{cphiw}
\end{equation}

We also need to compute the commutation relations with the electric
part of the hamiltonian, $E_{op}$, defined by

\begin{equation}
E_{op} = \sum_{\ell} E_{\ell}^{\ A} E_{\ell}^{\ A}
\end{equation}

\noindent They are

\begin{eqnarray}
\left[ E_{op}, \tilde{W} (L) \right] & = &
        \sum_{\ell, \ell '} \sum_{\in \ L}
                  \overline{\delta}_{\ell \ell '}
     \left\{ \tilde{W} (L_{xx}) \tilde{W} (L_{zz})
                       - \frac{1}{2} \tilde{W} (L) \right\}
        - 2 \sum_{\ell \in L}  \tilde{W} (\ell, \ L)
                                 \label{cEw}  \nonumber \\  \ \\
\left[ E_{op}, \phi (P_{x}^{\ y}) \right] & = &
     \sum_{\ell, \ell '} \sum_{\in \ P_{x}^{\ y}}
                              \overline{\delta}_{\ell {\ell '}}
     \left\{ \phi (P_{x}^{\ y}) \tilde{W} (L_{zz})
                     - \frac{1}{2} \phi(P_{x}^{\ y}) \right\}
    - 2 \sum_{\ell \in P_{x}^{\ y}}  \Pi (\ell , P_{x}^{\ y})
                                  \label{cEphi}   \nonumber \\ \ \\
\left[ E_{op}, \Gamma (P_{x}^{\ y}) \right] & = &
     \sum_{\ell, \ell '} \sum_{\in \ P_{x}^{\ y}}
                              \overline{\delta}_{\ell {\ell '}}
     \left\{ \Gamma (P_{x}^{\ y}) \tilde{W} (L_{zz})
                - \frac{1}{2} \Gamma (P_{x}^{\ y}) \right\}
    - 2 \sum_{\ell \in P_{x}^{\ y}}  \Lambda (\ell , P_{x}^{\ y})
                                \label{cEGamma}  \nonumber \\
\end{eqnarray}

\noindent where $z$ is the origin of the link $\ell$,

\begin{equation}
\overline{\delta}_{\ell {\ell '}}
           = \left\{ \begin{array}{rl}
                  1   & \mbox{for $\ell = \ell '$} \\
                 -1   & \mbox{for $\ell = \overline{\ell '}$} \\
                  0   & \mbox{otherwise}
                        \end{array}
                 \right.
\end{equation}

\noindent and $\tilde{W}(\ell ,L)$,
$\Pi(\ell , P_{x}^{\ y})$ and $\Lambda (\ell , P_{x}^{\ y})$
respectively are the gauge invariant conjugate operators of
$\tilde W(L)$, $\phi (P_{x}^{\ y})$ and $\Gamma (P_{x}^{\ y})$,

\begin{eqnarray}
\tilde{W}(\ell ,L) & = & Tr \left[\tilde{U}(L_{zz}) E_{\ell}
                                          \right]    \nonumber \\
\Pi(\ell , P_{x}^{\ y}) & = & \chi^{\dagger}_{_M} (x) \,
               \tilde{U}_{_{MN}} (P_{x}^{\ z}) \, X^{A}_{_{NS}}
    \tilde{U}_{_{ST}} (P_{z}^{\ y}) \, \chi_{_T} (y)
    \, E_{\ell}^{\ A}
                                                               \\
\Lambda(\ell , P_{x}^{\ y}) & = & \chi_{_M} (x)
   \varepsilon_{_{MN}} \, \tilde{U}_{_{NS}} (P_{x}^{\ z})
   \, X_{_{ST}}^{A}
    \tilde{U}_{_{TR}} (P_{z}^{\ y}) \, \chi_{_R}(y) \, E_{\ell}^{\ A}
                                                     \nonumber
\end{eqnarray}

\noindent where z is the origin of the link $\ell$ and x, y, z may
be even or odd. These operators are the analogous to the $T^{(1)}$
in the language of Rovelli-Smolin \cite{Rov88}.\\

Finally, in order to study the action of the path dependent
operators we need the commutators  \\

\begin{eqnarray}
\left[ \tilde{W}(L) , \tilde{W} (\ell, L') \right]
& = & \sum_{\ell ' \in L} \overline{\delta}_{\ell {\ell '}}
        \left\{ \tilde{W}({L'}_{zz} L_{zz})
       - \frac{1}{2} \tilde{W}(L) \tilde{W}(L')  \right\}
                                             \label{cwwl} \\
\left[ \tilde{W}(L) , \Pi_{0} (\ell , {P'}_{u}^{\ v}) \right]
& = & \sum_{\ell ' \in L}
                        \overline{\delta}_{\ell {\ell '}}
        \left\{ \phi_{0} ({P'}_{u}^{\ z} L_{zz} {P'}_{z}^{\ v})
             - \frac{1}{2} \phi_{0} ({P'}_{u}^{\ v}) \tilde{W}(L)
                                                \right\}
                          \label{cwpi} \nonumber \\ \ \\
\left[ \tilde{W}(L) , \Lambda_{2} (\ell , {P'}_{u}^{\ v}) \right]
& = & \sum_{\ell ' \in L}
                        \overline{\delta}_{\ell {\ell '}}
        \left\{ \Gamma_{2} ({P'}_{u}^{\ z} L_{zz} {P'}_{z}^{\ v})
             - \frac{1}{2} \Gamma_{2} ({P'}_{u}^{\ v}) \tilde{W}(L)
                                                \right\}
                          \label{cwlambda} \nonumber \\ \ \\
                                                   \nonumber \\
\left[ \phi_{0} (P_{x}^{\ y}) , \tilde{W} (\ell, L) \right]
& = & \sum_{\ell ' \in P_{x}^{\ y}}
                   \overline{\delta}_{\ell {\ell '}}
       \left\{ \phi_{0} (P_{x}^{\ z} L_{zz} P_{z}^{\ y})
           - \frac{1}{2} \phi_{0} (P_{x}^{\ y}) \tilde{W}(L_{zz})
                                    \right\} \nonumber \\ \ \\
                                                   \nonumber \\
\left[ \phi_{0} (P_{x}^{\ y}) , \Pi_{0} (\ell ,
                                         {P'}_{u}^{\ v}) \right]
& = & - \sum_{\ell ' \in P_{x}^{\ y}}
                           \overline{\delta}_{\ell {\ell '}}
    \left\{ \phi_{0} (P_{x}^{\ z} {P'}_{z}^{\ v})
                          \phi_{0} ({P'}_{u}^{\ z} P_{z}^{\ y})
                                          \right.  \nonumber  \\
&   & \hspace{3.5cm} + \left.
    \frac{1}{2} \phi_{0} (P_{x}^{\ y}) \phi_{0} ({P'}_{u}^{\ v})
                                     \right\} \label{cphipi} \\
                                                   \nonumber \\
\left[ \phi_{0} (P_{x}^{\ y}) , \Lambda_{2} (\ell ,
                                         {P'}_{u}^{\ v}) \right]
& = & - \sum_{\ell ' \in P_{x}^{\ y}}
                           \overline{\delta}_{\ell {\ell '}}
    \left\{ \Gamma_{2} ({P'}_{u}^{z} P_{z}^{\ y})
                   \phi_{0} (P_{x}^{\ z} {P'}_{z}^{\ v})
                                       \right.     \nonumber  \\
&   & \hspace{3.5cm} + \left.
 \frac{1}{2} \Gamma_{2} ({P'}_{u}^{\ v})  \phi_{0} (P_{x}^{\ y})
                                  \right\}  \label{cPhiLambda} \\
                                                   \nonumber \\
\left[ \Gamma_{2} (P_{x}^{\ y}) , \tilde{W} (\ell, L) \right]
& = & \sum_{\ell ' \in P_{x}^{\ y}}
                   \overline{\delta}_{\ell {\ell '}}
       \left\{ \Gamma_{2} (P_{x}^{\ z} L_{zz} P_{z}^{\ y})
           - \frac{1}{2} \Gamma_{2} (P_{x}^{\ y}) \tilde{W}(L_{zz})
                               \right\} \label{g2wl} \nonumber \\ \\
\left[ \Gamma_{2} ({P'}_{u}^{\ v}) , \Pi_{0} (\ell ,
                                         P_{x}^{\ y}) \right]
& = & - \sum_{\ell ' \in {P'}_{u}^{\ v}}
                           \overline{\delta}_{\ell {\ell '}}
   \left\{ \phi_{0} (P_{x}^{\ z} {P'}_{z}^{\ v})
                     \Gamma_{2} ({P'}_{u}^{\ z} P_{z}^{\ y})
                                        \right.    \nonumber  \\
&   & \hspace{3.5cm} + \left.
 \frac{1}{2} \phi_{0} (P_{x}^{\ y}) \Gamma_{2} ({P'}_{u}^{\ v})
                                   \right\}  \label{cGammapi} \\
\left[ \Gamma_{2} (P_{x}^{\ y}) , \Lambda_{2} (\ell ,
                                         {P'}_{u}^{\ v}) \right]
& = & - \sum_{\ell ' \in P_{x}^{\ y}}
                           \overline{\delta}_{\ell {\ell '}}
    \left\{ \Gamma_{2} (P_{x}^{\ z} {P'}_{z}^{\ v})
                            \Gamma_{2} ({P'}_{u}^{z} P_{z}^{\ y})
                                       \right.     \nonumber  \\
&   & \hspace{3.5cm} + \left.
 \frac{1}{2} \Gamma_{2} (P_{x}^{\ y}) \Gamma_{2} ({P'}_{u}^{\ v})
                                  \right\}  \label{cGammaLambda}
\end{eqnarray}

\noindent where $z$ is the origin of the link $\ell$ and $P_{x}^{\
z}$ is the portion of path $P_{x}^{\ y}$ from the site $x$ to the
site $z$. The algebra displayed in the Eqs.
(\ref{cphiphi}-\ref{cphiw}), (\ref{cEw}-\ref{cEGamma}) and
\mbox{(\ref{cwwl}-\ref{g2wl})} generalizes the algebra of gauge
invariant operators \cite{Gam86,Rov88} for the case of open paths.
With the help of this algebra we shall study the realization of the
operators in a Hilbert space spanned by the kets $\mid {\bf P}
\rangle$. \\

The action of the $\tilde{W}(L)$ operator is to add a loop to {\bf
P}

\begin{equation}
\tilde{W}(L) \mid {\bf P} \rangle = \mid {\bf P} + L \rangle
\end{equation}

\vspace{0.5cm}

The operators $\phi_0 (P_{x}^{\ y})$, $\Gamma_{2} (P_{x}^{\ y})$
and $\Gamma_{1}^{\dagger} (P_{x}^{\ y})$ add the open path
$P_{x}^{\ y}$ to the original collection of paths labeled by
{\bf P}

\begin{eqnarray}
\phi_{0} (P_{x_{e}}^{\ y_{o}}) \mid {\bf P} \rangle
              & = & \mid {\bf P} + P_{x_{e}}^{\ y_{o}} \rangle
                           \label{phi0I|>} \nonumber \\
\Gamma_{2} (P_{x_{o}}^{\ y_{o}}) \mid {\bf P} \rangle
             & = & \mid {\bf P} + P_{x_{o}}^{\ y_{o}} \rangle
                                   \label{Gamma2I|>} \\
\Gamma_{1}^{\dagger} (P_{x_{e}}^{\ y_{e}}) \mid {\bf P} \rangle
             & = & \mid {\bf P} + P_{x_{e}}^{\ y_{e}} \rangle
                           \label{Gamma1I|>}  \nonumber
\end{eqnarray}

The action of the operator $\phi_{3}$ on a mesonic path is to
annihilate the quarks of the ends of the path

\begin{eqnarray}
\phi_{3} ({P'}_{u}^{\ v}) \mid P_{x_{e}}^{\ y_{o}} \rangle
  & = & \delta_{xv} \delta_{yu}
        \tilde{W} ({P'}_{u_{o}}^{\ v_{e}} P_{x_{e}}^{\ y_{o}})
                                         \label{phi3I|>}
\end{eqnarray}

\noindent on a single barionic path is null

\begin{eqnarray}
\phi_{3} ({P'}_{u}^{\ v}) \mid P_{x_{o}}^{\ y_{o}} \rangle =
\phi_{3} ({P'}_{u}^{\ v}) \mid P_{x_{e}}^{\ y_{e}} \rangle = 0
\end{eqnarray}

\noindent and on two open paths is to joint the paths

\begin{eqnarray}
\lefteqn{\phi_{3} (P_{x}^{\ y})
   \mid {P'}_{u_{e}}^{\ v_{o}}, {P''}_{s_{e}}^{\ t_{o}} \rangle =}
                                                \nonumber \\
&& \delta_{xt} \, \delta_{yu}
      \mid P_{x_{o}}^{\ y_{e}} {P''}_{s_{e}}^{\ t_{o}},
                               {P'}_{u_{e}}^{\ v_{o}} \rangle
  - \delta_{xt} \delta_{yu}
        \mid {P''}_{s_{e}}^{\ t_{o}} P_{x_{o}}^{\ y_{e}}
                               {P'}_{u_{e}}^{\ v_{o}} \rangle
                                               \nonumber \\
&&  \hspace{6.5cm} - \delta_{xv} \delta_{ys}
         \mid {P'}_{u_{e}}^{\ v_{o}} P_{x_{o}}^{\ y_{e}}
                            {P''}_{s_{e}}^{\ t_{o}} \rangle
                                               \nonumber \\
\end{eqnarray}
\begin{eqnarray}
\lefteqn{\phi_{3} (P_{x}^{\ y})
   \mid {P'}_{u_{e}}^{\ v_{o}}, {P''}_{s_{o}}^{\ t_{o}} \rangle =}
                                                \nonumber \\
&& \delta_{yu} \, \delta_{xv}
   \mid P_{x_{o}}^{\ y_{e}} {P'}_{u_{e}}^{\ v_{o}},
                    {P''}_{s_{o}}^{\ t_{o}} \rangle
 - \delta_{yu} \, \delta_{xs}
   \mid {P'}_{v_{o}}^{\ u_{e}} P_{y_{e}}^{\ x_{o}}
                          {P''}_{s_{o}}^{\ t_{o}} \rangle
                                                \nonumber \\
&&  \hspace{6.5cm} - \delta_{yu} \, \delta_{xt}
   \mid {P'}_{v_{o}}^{\ u_{e}} P_{y_{e}}^{\ x_{o}}
                          {P''}_{t_{o}}^{\ s_{o}} \rangle
                                               \nonumber \\
\end{eqnarray}
\begin{eqnarray}
\lefteqn{\phi_{3} (P_{x}^{\ y})
    \mid {P'}_{u_{o}}^{\ v_{o}} {P''}_{s_{e}}^{\ t_{e}}, \rangle =}
                                                \nonumber \\
&& \delta_{xv} \, \delta_{yt}
   \mid {P''}_{s_{e}}^{\ t_{e}} P_{y_{e}}^{\ x_{o}}
                    {P'}_{v_{o}}^{\ u_{o}} \rangle
      + \delta_{xv} \, \delta_{ys}
   \mid {P''}_{t_{e}}^{\ s_{e}} P_{y_{e}}^{\ x_{o}}
                          {P'}_{v_{o}}^{\ u_{o}} \rangle
                                                \nonumber \\
&& \hspace{2.5cm} + \delta_{xu} \, \delta_{yt}
   \mid {P''}_{s_{e}}^{\ t_{e}} P_{y_{e}}^{\ x_{o}}
                    {P'}_{u_{o}}^{\ v_{o}} \rangle
      + \delta_{xu} \, \delta_{ys}
   \mid {P''}_{t_{e}}^{\ s_{e}} P_{y_{e}}^{\ x_{o}}
                          {P'}_{u_{o}}^{\ v_{o}} \rangle
                                               \nonumber \\
                                           \label{phi3F|>}
\end{eqnarray}

{}From the matricial element $\langle \Psi^{\dagger}_{u} , \Psi_{d}
, A_{\mu} \mid E_{\ell} \mid 0 \rangle = 0$ we see that the action
of the $E_{\ell}$ operator acting on the null path vanishes and,
using the conmutation relations of $E_{\ell}$ with path dependent
operators Eqs. (\ref{cEw}-\ref{cEGamma}), we may see that the
action of $E_{op}$ on the basis of path can be written as the sum
of four terms with a well defined geometrical action

\begin{eqnarray}
\lefteqn{E_{op} \mid {P_{1}}_{x_{1}}^{\ y_{1}} ,
                         {P_{2}}_{x_{2}}^{\ y_{2}} ,
        \ldots , {P_{m}}_{x_{m}}^{\ y_{m}} \rangle =}
                                           \nonumber \\
&& \left\{ E_{L} + E_{\Lambda} + E_{fus} + E_{fis} \right\}
      \mid {P_{1}}_{x_{1}}^{\ y_{1}} ,{P_{2}}_{x_{2}}^{\ y_{2}} ,
                 \ldots , {P_{m}}_{x_{m}}^{\ y_{m}} \rangle
                                          \label{Eop|>}
\end{eqnarray}

\noindent The first and second terms are diagonal

\begin{eqnarray}
E_{L} \mid {P_{1}}_{x_{1}}^{\ y_{1}} ,
                  \ldots , {P_{m}}_{x_{m}}^{\ y_{m}} \rangle
  & = & 2 \sum_{i=1}^{m} L_{i}
          \mid {P_{1}}_{x_{1}}^{\ y_{1}} ,
               \ldots , {P_{m}}_{x_{m}}^{\ y_{m}} \rangle \\
E_{\Lambda} \mid {P_{1}}_{x_{1}}^{\ y_{1}} ,
                  \ldots , {P_{m}}_{x_{m}}^{\ y_{m}} \rangle
   & = &  - \frac{1}{2} \sum_{i=1}^{m} \sum_{j=1}^{m}
                                              \Lambda_{ij}
               \mid {P_{1}}_{x_{1}}^{\ y_{1}} ,
                   \ldots , {P_{m}}_{x_{m}}^{\ y_{m}} \rangle
\end{eqnarray}

\noindent $L_{i}$ is the number of links of the ith path and

\begin{equation}
\Lambda_{ij} = \sum_{\ell \in P_{i}} \sum_{\ell ' \in P_{j}}
                                \overline{\delta}_{\ell \ell '}
\end{equation}

\noindent is a quadratic measure of the overlap between pairs of
paths in the list. It will be called the quadratic length. \\

The other two terms are related with fusion and fission effects on
paths. The fussion efects are produced by the presence of equal or
reversed links in different paths of the state [Fig. 1]. In
general it is given by

\begin{eqnarray}
\lefteqn{E_{fus} \mid {P_{1}}_{x_{1}}^{\ y_{1}} ,
       \ldots , {P_{m}}_{x_{m}}^{\ y_{m}} \rangle =}  \nonumber \\
& & - 2 {\sum_{i=1}^{m} \sum_{j=1}^{m}}_{i<j}
            \sum_{\ell \in P_{i}} \sum_{\ell ' \in P_{j}}
                                \overline{\delta}_{\ell \ell '}
  \mid {P_{1}}_{x_{1}}^{\ y_{1}} , \ldots ,
       {P_{i-1}}_{x_{i-1}}^{\ y_{i-1}} ,
             {P_{i}}_{x_{i}}^{\ z} {P_{j}}_{z}^{\ y_{j}},  \ldots
                                                     \nonumber \\
& & \hspace{4.5cm}
           \ldots , {P_{j-1}}_{x_{j-1}}^{\ y_{j-1}} ,
                  {P_{j}}_{x_{j}}^{\ z} {P_{i}}_{z}^{\ y_{i}},
                      \ldots , {P_{m}}_{x_{m}}^{\ y_{m}} \rangle
                                                 \label{EfusGral}
\end{eqnarray}

\noindent but this expression will depend on the kind of paths
involucrated in the fussion:

\begin{enumerate}

\item When the fusion is between two open paths the sum in $P_{i}$
and $P_{j}$ is given by

\begin{eqnarray}
\sum_{\ell \in P_{i}} \sum_{\ell ' \in P_{j}}
                           \overline{\delta}_{\ell \ell '}
  \mid {P_{1}}_{x_{1}}^{\ y_{1}} , & \ldots &,
     {P_{i-1}}_{x_{i-1}}^{\ y_{i-1}} ,
       {P_{i}}_{x_{i}}^{\ z} {P_{j}}_{z}^{\ y_{j}},  \ldots
                                                     \nonumber \\
    & \ldots & , {P_{j-1}}_{x_{j-1}}^{\ y_{j-1}} ,
                  {P_{j}}_{x_{j}}^{\ z} {P_{i}}_{z}^{\ y_{i}},
                      \ldots , {P_{m}}_{x_{m}}^{\ y_{m}} \rangle
\end{eqnarray}

\item If the fusion is between one loop ($P_{j} = L$) and an open
      path ($P_{i}$)

\begin{eqnarray}
- \sum_{\ell \in P_{i}} \sum_{\ell ' \in L}
                         \overline{\delta}_{\ell \ell '}
\mid {P_{1}}_{x_{1}}^{\ y_{1}} , & \ldots  &,
  {P_{i-1}}_{x_{i-1}}^{\ y_{i-1}} ,
    {P_{i}}_{x_{i}}^{z} L_{zz} {P_{i}}_{z}^{y_{i}}, \ldots
                                              \nonumber \\
     & \ldots &, {P_{j-1}}_{x_{j-1}}^{\ y_{j-1}},
           {P_{j+1}}_{x_{j+1}}^{\ y_{j+1}}  \ldots ,
             {P_{m}}_{x_{m}}^{\ y_{m}} \rangle
\end{eqnarray}

\item When the fusion is between two loops, $L_{i}$ and $L_{j}$, we
      have that

\begin{eqnarray}
- \sum_{\ell \in L_{i}} \sum_{\ell ' \in L_{j}}
                         \overline{\delta}_{\ell \ell '}
\mid {P_{1}}_{x_{1}}^{\ y_{1}} , & \ldots &
    {P_{i-1}}_{x_{i-1}}^{\ y_{i-1}} ,
      {L_{i}}_{zz} {L_{j}}_{zz} , \ldots
                                              \nonumber \\
      & \ldots &,  {P_{j-1}}_{x_{j-1}}^{\ y_{j-1}},
           {P_{j+1}}_{x_{j+1}}^{\ y_{j+1}}  \ldots ,
             {P_{m}}_{x_{m}}^{\ y_{m}} \rangle
                                         \label{EfusLL}
\end{eqnarray}

\end{enumerate}

\noindent In Eqs. (\ref{EfusGral}-\ref{EfusLL}) the site $z$ is the
origin of the link $\ell$. Therefore, the fussion term creates one
new path from the product of colliding paths. Decreasing the path
number in the state by one. \\

Finally, the fission term contributes when there are multiple links
inside one path, [Fig. 2]. In these cases the paths breaks into two
parts increasing the number of paths by one. The action of
$E_{fis}$ is given by

\begin{eqnarray}
E_{fis} \mid {P_{1}}_{x_{1}}^{\ y_{1}} ,
                  \ldots , {P_{m}}_{x_{m}}^{\ y_{m}} \rangle =
    \sum_{i=1}^{m} \sum_{\ell , \ell ' \in P_{i}}
                              \overline{\delta}_{\ell \ell '}
     \mid {P_{1}}_{x_{1}}^{\ y_{1}} , \ldots
          , {P_{i}}_{x_{i}}^{\ z} {P_{i}}_{z}^{\ y_{i}}, L_{zz},
                     \ldots , {P_{m}}_{x_{m}}^{\ y_{m}} \rangle
                                     \label{Efis|>}   \nonumber \\
\end{eqnarray}

\noindent where z is the origin of the link $\ell$. \\


\section{Mandelstam Identities for Open Paths}
\label{sec:Man-Id}

It is important to realize that not all states $\mid {\bf P}
\rangle$ are independent because the SU(2) group properties and the
{\bf Mandelstam identities of the first and second kind}
\cite{Man79,Gil81,Gam86,Gam89,Lol92a,Lol93a}:

\begin{eqnarray}
\tilde{W}(L_{1} \circ L_{2}) & = & \tilde{W}(L_{2} \circ L_{1})
                                                  \label{MI1} \\
\tilde{W}(L_{1}) \, \tilde{W}(L_{2}) & = &
                \tilde{W}(L_{1} \circ L_{2}) + \tilde{W}(L_{1}
                          \circ \overline{L_{2}})  \label{MI2}
\end{eqnarray}

\noindent impose relations among them [Fig. 3a]. \\

The usual Mandelstam identities connect only loops, we have now
new identities that connect open paths and loops

\begin{eqnarray}
\tilde{W}(L) \, \phi_{0} (P_{x}^{\ y})
      & = & \phi_{0} (P_{x}^{\ z} L_{zz} P_{z}^{\ y})
                  + \phi_{0} (P_{x}^{\ z}
                  \overline{L}_{zz} P_{z}^{\ y})  \label{MI3a} \\
                                               \nonumber \\
\tilde{W}(L) \, \Gamma_{2} (P_{x}^{\ y})
      & = & \Gamma_{2} (P_{x}^{\ z} L_{zz} P_{z}^{\ y})
                  + \Gamma_{2} (P_{x}^{\ z}
                  \overline{L}_{zz} P_{z}^{\ y})
                                            \label{MI3b} \\
                                               \nonumber \\
\tilde{W}(L) \, \Gamma_{1}^{\dagger} (P_{x}^{\ y})
      & = & \Gamma_{1}^{\dagger} (P_{x}^{\ z} L_{zz} P_{z}^{\ y})
                  + \Gamma_{1}^{\dagger} (P_{x}^{\ z}
                  \overline{L}_{zz} P_{z}^{\ y})
                                                \label{MI3c}
\end{eqnarray}

\noindent where $z$ is a contact point between the loop $L$ and the
open paths $P$ \mbox{[Fig. 3b]}. It is important to notice that is
not necessary that the loop $L$ and the open path have a common
point since using the cyclic property of the loop functional
$\tilde{W}$ we always may choose a loop

$$ L' = Q_{z}^{\ w} \, L_{ww} \, Q_{w}^{\ z} $$

\noindent where $z$ is a point of the open path and, for instance,
the Eq. (\ref{MI3a}) may be rewritten as

\begin{eqnarray}
\tilde{W} (L) \, \phi_{0} (P_{x}^{\ y})
& = &  \tilde{W} (L') \, \phi_{0} (P_{x}^{\ y})    \nonumber  \\
& = &
\phi_{0} (P_{x}^{\ z} Q_{z}^{\ w} L_{ww} Q_{w}^{\ z} P_{z}^{\ y})
    + \phi_{0} (P_{x}^{\ z} Q_{z}^{\ w}
            \overline{L}_{ww} Q_{w}^{\ z} P_{z}^{\ y})
                                               \nonumber  \\
&   &
\end{eqnarray}

\noindent This identity holds for any path $Q$. \\

For $L_{1} = L_{0}$ (the null loop) we have from the Mandelstam
identity of first kind, Eq. (\ref{MI1}), that

\begin{equation}
\tilde{W}(L) = \tilde{W}(\overline{L})  \label{MI2'}
\end{equation}

\noindent that is, the global orientation of the loop does not
matter. \\

We have also a new type of Mandelstam indentities that exclusively
connect operators of open paths, this constraints involve the end
points of the paths connecting mesonic and barionic operators. The
fermionic character of the $\chi$ fields ensure that

\begin{equation}
\chi_{_A}(y) \, \chi_{_B}(y) = \textstyle{\frac{1}{2}} \left\{
 \chi_{_A}(y) \, \chi_{_B}(y) - \chi_{_B}(y) \,
                                         \chi_{_A}(y) \right\}
    = \textstyle{\frac{1}{2}} \, \varepsilon_{_{AB}} \,
              \varepsilon_{_{ST}} \chi_{_S}(y) \, \chi_{_T}(y)
\end{equation}

\noindent then, for two open paths with the same end points we have
[Fig. 4]

\begin{eqnarray}
\phi_{0}(P_{x}^{\ y}) \, \phi_{0}({P'}_{x}^{\ y})
  & = &  \frac{1}{4}
         \tilde{W}(P_{y}^{\ x} {P'}_{x}^{\ y}) \,
           \Gamma_{1}^{\dagger}(P_{o}^{xx}) \,
           \Gamma_{2}(P_{o}^{yy})   \label{PI1}     \\
\Gamma_{2} (P_{x}^{\ y}) \,
                        \Gamma_{2}({P'}_{x}^{\ y})
  & = & - \frac{1}{4} \,
          \tilde{W}({P'}_{x}^{\ y} P_{y}^{\ x})
          \, \Gamma_{2}(P_{o}^{xx}) \,
             \Gamma_{2}(P_{o}^{yy})   \label{PI2}   \\
\Gamma_{1}^{\dagger} (P_{x}^{\ y}) \,
                    \Gamma_{1}^{\dagger} ({P'}_{x}^{\ y})
  & = & - \frac{1}{4} \,
          \tilde{W}({P'}_{x}^{\ y} P_{y}^{\ x})
          \, \Gamma_{1}^{\dagger}(P_{o}^{xx}) \,
             \Gamma_{1}^{\dagger}(P_{o}^{yy})   \label{PI2b}
\end{eqnarray}

\noindent These identities together with the Mandelstam constraints
(\ref{MI2}) allow to obtain the identities (\ref{MI3a}-\ref{MI3c}).
For two open paths with one equal end point \mbox{[Fig. 5]} we have

\begin{eqnarray}
\phi_{0}(P_{x}^{\ y}) \, \phi_{0}({P'}_{u}^{\ y})
  & = &  \frac{1}{2}
 \Gamma_{1}^{\dagger}({P'}_{u}^{\ y} P_{y}^{\ x})
                     \, \Gamma_{2}(P_{o}^{yy}) \label{PI3} \\
\phi_{0}(P_{x}^{\ y}) \, \phi_{0}({P'}_{x}^{\ v})
  & = &  \frac{1}{2}
         \Gamma_{1}^{\dagger}(P_{o}^{xx}) \,
         \Gamma_{2}(P_{y}^{\ x} {P'}_{x}^{\ v}) \label{PI4} \\
\phi_{0}(P_{x}^{\ y}) \, \Gamma_{2}({P'}_{u}^{\ y})
  & = &  - \frac{1}{2}
  \phi_{0}(P_{x}^{\ y} {P'}_{y}^{\ u})
         \, \Gamma_{2}(P_{o}^{yy})    \label{PI5}   \\
\phi_{0} (P_{x}^{\ y}) \,
                  \Gamma_{1}^{\dagger}({P'}_{x}^{\ v})
  & = & - \frac{1}{2}  \Gamma_{1}^{\dagger}(P_{o}^{xx}) \,
        \phi_{0}({P'}_{v}^{\ x} P_{x}^{\ y}) \label{PI6}  \\
\Gamma_{2} (P_{x}^{\ y}) \,
          \Gamma_{2}({P'}_{u}^{\ y})
  & = & - \frac{1}{2} \,
          \Gamma_{2}(P_{x}^{\ y} {P'}_{y}^{\ u})
                  \Gamma_{2}(P_{o}^{yy}) \label{PI7}  \\
\Gamma_{1}^{\dagger} (P_{x}^{\ y}) \,
                 \Gamma_{1}^{\dagger} ( {P'}_{u}^{\ y})
  & = & - \frac{1}{2} \,
          \Gamma_{1}^{\dagger} (P_{x}^{\ y} {P'}_{y}^{\ u})
          \, \Gamma_{1}^{\dagger} (P_{o}^{yy}) \label{PI8}
\end{eqnarray}

\noindent where $P_{o}$ is the null path. \\

Finally, we can see that the product
$\chi_{_A} (y_{o}) \, \chi_{_B} (y_{o}) \, \chi_{_C} (y_{o})$
is null it since will have two identical indices. Therefore, two is
the maximum number of quarks that we can put in a site, then,
we have

\begin{eqnarray}
\phi_{0}(P_{x}^{\ y}) \, \Gamma_{2}({P'}_{y}^{\ y}) & = &
\phi_{0}(P_{x}^{\ y}) \, \Gamma_{1}^{\dagger}({P'}_{x}^{\ x}) = 0
                              \nonumber \\       \label{PI9}  \\
\Gamma_{2}(P_{x}^{\ y}) \, \Gamma_{2}({P'}_{y}^{\ y}) & = &
\Gamma_{1}^{\dagger}(P_{x}^{\ y}) \,
       \Gamma_{1}^{\dagger}({P'}_{x}^{\ x}) = 0
                                                    \nonumber
\end{eqnarray}

\noindent and

\begin{eqnarray}
\phi_{0}(P_{x}^{\ y}) \, \phi_{0}({P'}_{u}^{\ y}) \,
                         \phi_{0}({P''}_{s}^{\ y})  & = &
\Gamma_{2}(P_{x}^{\ y}) \, \Gamma_{2}({P'}_{u}^{\ y}) \,
                 \Gamma_{2}({P''}_{s}^{\ y})  = 0  \label{PI10}
\end{eqnarray}

\noindent which are not independent constraints because they may be
obtained from the constraints (\ref{PI3}) and (\ref{PI7}) with the
(\ref{PI9}). \\

The Mandelstam identities may be used to reduce the redundancy of
the basis of open paths and loops. For instance, if we consider a
loop $L$ containing a double link $\ell$, we may write $L$ as $A
\ell B \ell$ with A and B closed parts of $L$, [Fig. 5a]. Then,
from the Mandelstam identity of second kind, Eq. (\ref{MI2}), we
have

\begin{equation}
 \tilde{W}(L) = \tilde{W}(L_{1} L_{2})
                       = \tilde{W}(A \ell B \ell ) =
  \tilde{W}(A \ell ) \, \tilde{W}(B \ell )
                        - \tilde{W}(A \overline{B})  \label{MI2b}
\end{equation}

\noindent therefore, a state containing loops with multiple links
may be expressed as a linear combination of states where the links
have reduced multiplicity. Then, in forming the basis, one has only
to consider states with loops without multiple links. From the
identities (\ref{MI3a}-\ref{MI3c}) we can obtain a similar
expression for open paths with a double link [Fig. 5b]. Finally,
the constraints for open paths (\ref{PI1}-\ref{PI10}) allow one to
work with a basis where different open paths do not have the same
end points.


\section{Hamiltonian Dynamics}
\label{sec:Ham-Dyn}

The hamiltonian is given by

\begin{equation}
H = \frac{g^{2}}{2} \left( W_{E} + W_{m} + \lambda W_{q} +
                         \lambda ^{2} W_{M} \right) \label{Ham}
\end{equation}

\noindent with $\lambda = 1/g^{2}$ and

\begin{eqnarray}
W_{E} & = & E_{op} \\
W_{m} & = & m \sum_{\vec{r}} (-1)^{x+y+z} \chi^{\dagger} (\vec{r})
                                              \chi (\vec{r}) \\
W_{q} & = & \sum_{\vec{r} , \widehat{n}}
            \chi^{\dagger} (\vec{r}) \tilde{U}(\vec{r}, \widehat{n})
                  \chi (\vec{r} + \widehat{n}) + H. c. \\
W_{M} & = & - \sum_{\Box} \left( W(\Box) + W^{\dagger} (\Box) \right)
\end{eqnarray}

\noindent where $\vec{r} = (x,y,z)$ labels the sites and $\Box$ the
plaquettes \cite{Ban77}. \\

The action of the electric term, $W_{E}$, is given by the action of
the electric operator $E_{op}$, Eqs. (\ref{Eop|>}-\ref{Efis|>}).
Then, $W_{E}$ will give a measure of the length of the paths, the
overlap between couples of paths (quadratic length) and the
interaction effects among paths (fission and fussion terms). \\

We can see that the action of the mass term, $W_{m}$, over a state
$\mid {\bf P} \rangle$ is given by

\begin{equation}
W_{m} \mid {\bf P} \rangle =
 m \left( 2 \, {\cal N} ({\bf P}) - \frac{1}{2} N_{s} \right)
                                \mid {\bf P} \rangle
\end{equation}

\noindent where ${\cal N} ({\bf P})$ is the number of open paths in
the state $\mid {\bf P} \rangle$ and $N_{s}$ is the number of
lattice sites. The action of $W_{m}$ justifies our election of the
even-odd orientation of the mesonic paths since for a state
$\mid P_{u_{e}}^{\ v_{o}} \rangle$ we have

\begin{equation}
W_{m} \mid P_{u_{e}}^{\ v_{o}} \rangle
              = m \left( 2 - \frac{1}{2} N_{s} \right)
                              \mid P_{u_{e}}^{\ v_{o}} \rangle
\end{equation}

\noindent while if we choose the other representation, where we
interchange the parity of the sites, we have

\begin{equation}
W_{m} \mid P_{u_{o}}^{\ v_{e}} \rangle
  = - m \left(2 - \frac{1}{2} N_{s} \right)
                       \mid P_{u_{o}}^{\ v_{e}} \rangle
\end{equation}

\noindent Then, the first representation (where the even sites are
the starting points of the paths) has less energy and is the
natural choice for the orientation of the mesonic paths. Thus, as
it is well known, the mass term breaks the chiral symmetry. \\

The interaction term, $W_{q}$ , can be written in terms of the
operators $\phi_{o}$ and $\phi_{3}$ as

\vspace{0.5cm}

\begin{equation}
W_{q} = \sum_{\vec{r}_{e},\widehat{n}} \phi_{0} (\ell_{r})
    + \sum_{\vec{r}_{o},\widehat{n}} \phi_{3} (\ell_{r})  + H. c.
\end{equation}

\noindent where the sum is over the six directions $\widehat{n}$;
and $\ell_{r}$ is the link starting in $\vec{r}$ and ending in
$\vec{r} + a \widehat{n}$, with $a$ equal to the lattice spacing.
Therefore, through the action of $\phi_{0}$ the term $W_{q}$ can
create mesonic and barionic states. If we compute the perturbation
expansion in $\lambda = 1/g^{2}$ in the strong-coupling region
for the rescaled hamiltonian

\vspace{0.5cm}

$$
W = W_{E} + W_{m} + \lambda \, W_{q} + \lambda^{2} \, W_{M}
$$

\noindent we can see that the mesonic states arise in the first
order of the expansion and the barionic states in the second order.
{}From the action of $\phi_{0}$, Eq. (\ref{phi0I|>}), the term
$W_{q}$ produces mesonic states when perturbs the zeroth order of
the expansion

\vspace{0.5cm}

$$
W_{q} \mid 0 \rangle =
   \sum_{\vec{r}_{e}, \widehat{n}} \phi_{0} (\ell_{r})
                                        \mid 0 \rangle =
   \sum_{\vec{r}_{e}, \widehat{n}} \mid \ell_{r} \rangle
$$

\noindent and, from the Mandelstam constraints (\ref{PI1}),
(\ref{PI3}) and (\ref{PI4}), $W_{q}$ generates barionic states when
acts over the mesonic states of the first order

\vspace{1cm}

\begin{eqnarray}
W_{q} \mid \ell_{\vec{r}}^{\, \vec{r}+\widehat{n}} \rangle
 & = &
   \sum_{\vec{s}_{e}, \widehat{m}}
      \phi_{0} ({\ell '}_{\vec{s}}^{\, \vec{s}+\widehat{m}})
        \mid \ell_{\vec{r}}^{\, \vec{r}+\widehat{n}} \rangle
   =
   \sum_{\vec{s}_{e}, \widehat{m}}
      \phi_{0} ({\ell '}_{\vec{s}}^{\, \vec{s}+\widehat{m}})
        \, \phi_{0} (\ell_{\vec{r}}^{\, \vec{r}+\widehat{n}})
                                             \mid 0 \rangle
                                                \nonumber \\
 & = &
  \sum_{\vec{s}_{e} \neq \vec{r}_{e}, \widehat{m} \neq \widehat{n}}
      \phi_{0} ({\ell '}_{\vec{s}}^{\, \vec{s}+\widehat{m}})
        \, \phi_{0} (\ell_{\vec{r}}^{\, \vec{r}+\widehat{n}})
                                             \mid 0 \rangle
                                                \nonumber \\
&   & \hspace{1.5cm}
  + \frac{1}{2} \sum_{\widehat{m} \neq \widehat{n}}
     \Gamma_{1}^{\dagger} (P_{0}^{\vec{r} \vec{r}}) \,
       \Gamma_{2} (\ell_{\vec{r}+\widehat{n}}^{\, \vec{r}}
                {\ell '}_{\vec{r}}^{\, \vec{r}+\widehat{m}})
                                             \mid 0 \rangle
                                                \nonumber \\
&   & \hspace{3.0cm}
+ \frac{1}{4} \tilde{W} (\ell_{\vec{r}+\widehat{n}}^{\, \vec{r}}
                      {\ell '}_{\vec{r}}^{\, \vec{r}+\widehat{n}}) \,
    \Gamma_{1}^{\dagger} (P_{0}^{\vec{r} \vec{r}}) \,
    \Gamma_{2} (P_{0}^{\vec{r}+\widehat{n},\vec{r}+\widehat{n}})
                                             \mid 0 \rangle
                                                \nonumber
\end{eqnarray}

\noindent where $P_{0}$ is the null path.
Finally, through the action
of $\phi_{3}$ the term $W_{q}$ can also joint two open paths, Eqs.
(\ref{phi3I|>}-\ref{phi3F|>}). \\

\noindent In according to (\ref{WW}) and (\ref{MI2'}) we can write
the magnetic term as

\begin{equation}
W_{M} = \sum_{\Box} \left( \tilde{W}(\Box)
                   + \tilde{W}^{\dagger}(\Box) \right)
  = 2 \sum_{\Box} \tilde{W}(\Box)
              = 4 \sum_{\Box > 0} \tilde{W}(\Box)
\end{equation}

\noindent then, the action of $W_{M}$ is given by

\begin{equation}
W_{M} \mid P \rangle = 4 \sum_{\Box > 0} \mid P \Box \rangle
\end{equation}

\noindent Now, we shall examine the symmetries of the hamiltonian
(\ref{Ham}) in this representation, in each case we describe the
transformation laws of the gauge invariant operators. As we are
going to show latter on, these six symmetries allow one to use the
invariance of the vacuum state in order to reduce the redundancy of
the basis elements. \\

\begin{description}
\item [Lattice translation by even number of links ] \  \\
        The parity of the sites remain invariant under
        translation by an even number of links
        \begin{equation}
        \chi (x,y,z) \longrightarrow
             \chi (x + 2 n_{x}, \, y + 2 n_{y}, \, z + 2 n_{z})
        \end{equation}
        where $\ell$, $m$ and $n$ are integer numbers. The path
        dependent operators transform as
        \begin{eqnarray}
        \tilde{W}(L) & \longrightarrow & \tilde{W}(L')
                                                    \nonumber \\
        \phi ( P_{r}^{\ s} ) & \longrightarrow &
              \phi ( P_{r+2 \hat{n}}^{\ s+2 \hat{n}})   \\
        \Gamma_{2} ( P_{r}^{\ s}) & \longrightarrow &
              \Gamma_{2} (P_{r+2\hat{n}}^{\ s+2\hat{n}})
                                                    \nonumber \\
        \Gamma_{1}^{\dagger} (P_{r}^{\ s}) & \longrightarrow &
            \Gamma_{1}^{\dagger} (P_{r+2\hat{n}}^{\ s+2\hat{n}})
                                                    \nonumber
        \end{eqnarray}
        where $2 \, \hat{n} = (2\,n_{x}, 2\,n_{y}, 2\,n_{z})$
        and $L'$ is the traslated loop.

\item [Lattice translation by odd number of links ] \  \\
        For translations of a link in the $\widehat{z}$ direction
        the even sites transform into odd sites and vice versa, and
        the operators transform as
        \begin{eqnarray}
        \tilde{W}(L) & \longrightarrow & \tilde{W}(L')
                                                \nonumber \\
        \chi (\vec{r}) & \longrightarrow & -i (-1)^{x_{r}}
                           \chi (\vec{r} + \hat{n}_{z})
                                                \nonumber \\
           \phi \left(P_{r}^{\ s}\right) & \longrightarrow &
            \phi^{\dagger} \left(P_{s+n_{z}}^{\ r+n_{z}} \right)
                                                  \label{Odd} \\
        \Gamma_{2} (P_{r}^{\ s}) & \longrightarrow &
           - \Gamma_{1} (P_{r+n_{z}}^{\ s+n_{z}})
                                                    \nonumber \\
        \Gamma_{1}^{\dagger} (P_{r}^{\ s}) & \longrightarrow &
           - \Gamma_{2}^{\dagger} (P_{r+n_{z}}^{\ s+n_{z}})
                                                    \nonumber
        \end{eqnarray}

\item [Rotations by $\pi /2$ ] \  \\
         For rotations around direction $\widehat{n}_{z}$ we have
         \begin{eqnarray}
           \tilde{W}(L)       & \longrightarrow &
                                           T(L) \  \tilde{W}(L')
                                                     \nonumber \\
           \chi(\vec{r})      & \longrightarrow &
                                       R(x,y,z) \chi (\vec{r}\,')
                                                     \nonumber \\
           \phi (P_{r}^{\ s}) & \longrightarrow &
                  R(\vec{r}) \, R(\vec{s}) \,
                        T(P_{r}^{\ s}) \ \phi (P_{r'}^{\ s'}) \\
        \Gamma_{2} (P_{r}^{\ s}) & \longrightarrow &
              R(\vec{r}) \, R(\vec{s}) \,
                    T(P_{r}^{\ s}) \ \Gamma_{2} (P_{r'}^{\ s'})
                                                    \nonumber \\
        \Gamma_{1}^{\dagger} (P_{r}^{\ s}) & \longrightarrow &
          R(\vec{r}) \, R(\vec{s}) \,
           T(P_{r}^{\ s}) \ \Gamma_{1}^{\dagger} (P_{r'}^{\ s'})
                                                    \nonumber
        \end{eqnarray}
        where
        \begin{equation}
        R(\vec{r}) \equiv \frac{1}{2} \left( (-1)^{x_{r}}
          + (-1)^{y_{r}} + (-1)^{z_{r}} - (-1)^{x_{r}+y_{r}+z_{r}}
                                                        \right)
        \end{equation}
        and
        \begin{equation}
        T(P) \equiv
         \left( \prod_{\ell_{x} \in P} (-1)^{y_{\ell}+z_{\ell}}
                                                          \right)
         \left( \prod_{\ell_{y} \in P} (-1)^{z_{\ell}+x_{\ell}}
                                                          \right)
         \left( \prod_{\ell_{z} \in P} (-1)^{x_{\ell}+y_{\ell}}
                                                          \right)
        \end{equation}
        For the case of a plaquette and a link we obtain
        \begin{eqnarray}
              \tilde{W}(\Box) & \longrightarrow &
                                  \tilde{W}(\Box ') \nonumber \\
              \phi (\ell_{x}) & \longrightarrow &
                         \phi (\ell_{y})            \nonumber \\
              \phi (\ell_{y}) & \longrightarrow &
                         \phi (\ell_{-x})         \label{R90} \\
              \phi (\ell_{z}) & \longrightarrow &
                               \phi (\ell_{z})      \nonumber
        \end{eqnarray}
        and for path with two links
        \begin{eqnarray}
            \Gamma_{2} (P_{r}^{\ s}) & \longrightarrow &
                      \Gamma_{2} (P_{r'}^{\ s'})     \nonumber \\
            \Gamma_{1}^{\dagger} (P_{r}^{\ s}) & \longrightarrow &
                \Gamma_{1}^{\dagger} (P_{r'}^{\ s'})
        \end{eqnarray}

\item [Rotations by $\pi$ about a lattice center ]  \   \\
         These are the rotations around any of the three axes
         passing through the geometrical center of a lattice cube
          \begin{eqnarray}
              \tilde{W}(L) & \longrightarrow & \tilde{W}(L')
                                                     \nonumber  \\
              \chi (\vec{r}) & \longrightarrow & \chi (\vec{r'})
                                                    \nonumber \\
              \phi (P_{r}^{\ s}) & \longrightarrow &
                                        \phi (P_{r'}^{\ s'}) \\
        \Gamma_{2} (P_{r}^{\ s}) & \longrightarrow &
                      \Gamma_{2} (P_{r'}^{\ s'})
                                                    \nonumber \\
        \Gamma_{1}^{\dagger} (P_{r}^{\ s}) & \longrightarrow &
                       \Gamma_{1}^{\dagger} (P_{r'}^{\ s'})
                                                    \nonumber
          \end{eqnarray}

\item [ Parity ]  \ \\
          This is the reflection through the origin.
          \begin{eqnarray}
            \tilde{W}(L) & \longrightarrow & \tilde{W}(L')
                                                \nonumber  \\
            \chi (\vec{r}) & \longrightarrow & \chi (- \vec{r} )
                                                \nonumber  \\
            \phi (P_{r}^{\ s}) & \longrightarrow &
               \phi (P_{-r}^{\ -s})           \\
        \Gamma_{2} (P_{r}^{\ s}) & \longrightarrow &
                   \Gamma_{2} (P_{-r}^{\ -s})
                                                    \nonumber \\
        \Gamma_{1}^{\dagger} (P_{r}^{\ s}) & \longrightarrow &
                   \Gamma_{1}^{\dagger} (P_{-r}^{\ -s})
                                                    \nonumber
         \end{eqnarray}
         \\ \ \\ \ \\
\item [G Parity ] \  \\
          This is the complex conjugation of the operators
          \begin{eqnarray}
              \tilde{W}(L) & \longrightarrow & \tilde{W}^{\dagger}(L)
                    = \tilde{W}(\overline{L}) = \tilde{W}(L)
                                                   \nonumber \\
              \phi_{0} (P_{r}^{\ s}) & \longrightarrow &
                      \phi_{0}^{\dagger} (P_{r}^{\ s})
                                     = \phi_{3} (P_{s}^{\ r})
                                                   \nonumber \\
              \phi_{3} (P_{r}^{\ s}) & \longrightarrow &
                    \phi_{3}^{\dagger} (P_{r}^{\ s})
                         = \phi_{0} (P_{s}^{\ r})   \label{Gpar} \\
        \Gamma_{2} (P_{r}^{\ s}) & \longrightarrow &
           \Gamma_{2}^{\dagger} (P_{r}^{\ s})
                                                    \nonumber \\
        \Gamma_{1}^{\dagger} (P_{r}^{\ s}) & \longrightarrow &
           \Gamma_{1} (P_{r}^{\ s})
                                                    \nonumber
          \end{eqnarray}
\end{description}

\noindent The identification of these symmetries in the continuum
was studied by Banks, Kogut and Susskind in \cite{Ban77,Sus77}. \\

This path-dependent formalism allows to accomplish explicit
calculations of the vacuum energy density and the mass-gap solving
the Schr\"{o}dinger equation by means of cluster approximation
\cite{Gam89}. The idea of this approximation is to consider a
Hilbert space restricted to a subspace of the entire space of paths
in the lattice. This space is spanned by a finite basis of simple
paths, without repeated links, or clusters. We define a cluster as
a list of open paths and loops confined in a finite spatial region,
under this approximation a generic path in the lattice is
understood as an unordered list of clusters separated by formally
infinite interdistances. Then, a physical state is described,
specifying the occupation number of each nonequivalent cluster, as
$\mid n_{1}, n_{2}, \ldots , n_{m}, \ldots \rangle$ where $n_{i}$
denotes the number of times that cluster $i$ appears in the list of
paths that describes the state. \\

We are mainly interested in a description of the quantum ground
state of the gauge system. The invariance of this state under the
symmetries of the hamiltonian allows one to eliminate equivalent
elements of the path basis. For instance, by application of the
symmetries (\ref{Odd}), (\ref{R90}) and (\ref{Gpar})
we can show that all the
mesonic states with only one link are equivalent, independently of
its origin or direction, the same occurs for the plaquettes. \\

We need to work with a finite number of clusters, to do so that we
introduce an order notion between clusters. This may be done in a
recursive way, the null path is zeroth order, the path of one link
and one plaquette constitute the first order, The Nth order will
include all the paths that are obtained from the \mbox{(N-1)th}
order by addition of plaquettes or links, through application of
the terms $\tilde{W}$ and $\phi_{0}$ of the hamiltonian.
The idea is to give a set of rules in order to build any element of
the basis. To do that we apply the operators $\phi_{0}$ and
$\tilde{W}$ as follows: we only compute the action of the operator
$\phi_{0} (\ell)$ over a loop $L$ when $\ell$ is a link of $L$, and
the action of the operator $\tilde{W}(\Box)$ over open paths and
loops when some of the four links of the plaquette coincides with
some of the path links. The action of $\phi_{0} (\ell)$ over an
open path is to add links in the neighborhood of the path. All
other links and plaquettes do not generate other clusters, they
only increase the occupation numbers of clusters associates to one
link and one plaquette. \\

Independent elements of the basis are obtained when the electric
operator and $\phi_{3}$ act on the given elements of order N. The
operator $\phi_{3} (\ell)$ joins two open paths that are separated
by one link. The electric part of the hamiltonian will also
generate clusters of the Nth order through the fusion and fission
terms. The fusion produce clusters with multiple links and the
fission term breaks loops and produces disconnected clusters.
Finally, the Mandelstam constraints for open paths, Eqs.
(\ref{PI1}), (\ref{PI3}) and (\ref{PI4}), allow one to obtain the
barionic states. \\

In the generation of the clusters we must check if the new cluster
is not equivalent, under some of the hamiltonian symmetries, to any
one of the already generated clusters or, through Mandelstam
identities for loops, Eqs. (\ref{MI1}-\ref{MI2}), and the new
Mandelstam constraints for open paths, Eqs.
(\ref{MI3a}-\ref{MI3c}), to a combination of the already generates
clusters. \\

The basis generated must be truncated for computational purposes.
The ge\-ne\-ra\-tion is iterated up to a maximum order $N_{max}$.
In \mbox{Fig. 6} we show the basis generated with this rules for
the order $N_{max} = 2$ (we have chosen the symbol $\bullet$ to
represent one antiquark and a $\circ$ to represent one quark). \\

It is well known from previous applications
\cite{Gam89,Aro,For91,For92a,DiB} that the cluster approximation
works quite well in the \mbox{strong-coupling} region, and by
considering a big enough basis of clusters and introducing some
collective variables it is possible to reach the
\mbox{weak-coupling} regime. \\

In a following paper we will present the results obtained using
this approximation, for the quantum ground state energy and the
mass-gap for the SU(2) QCD .


\section{Conclusion}
\label{sec:Con}

The hamiltonian formalism of lattice QCD in the path representation
has been introduced. Each term of the hamiltonian has a well
defined geometrical effect inducing deformations or rearrangements
over paths representing the physical states. We have restricted
ourselves in this paper to the introduction of the general
formalism and the study of the Mandelstam identities for the
$SU(2)$ QCD. Even though it is straightforward to extend this
formalism to full QCD, there is a renewed interest in the study of
the $SU(2)$ gauge theory with dynamical fermions \cite{Dag91}.
Besides reasons of economy in computational time
\cite{Mar81,Bil86,Wei81,Lac87} this study is also motivated by some
recent developments in the area of High-$T_{c}$ superconductivity
leading to a Heisenberg model written as a fermionic theory. A
connection between this hamiltonian and a $SU(2)$ lattice gauge
theory with dynamical fermions in $2+1$ dimensions has been
recently made \cite{Dag91}. Dagotto, Koci\'{c} and Kogut have
pointed out that a manifestly gauge invariant approach should be
developed to guide the computational work in this problem. We hope
that the present formulation could help to treat this problem. \\

We wish to thank D. Armand-Ugon, J. M. Aroca, H. Fort, H. Morales
and R. Siri for valuable discussions and comments. The work of
\mbox{L. S.} was supported, in part, by a grant of Consejo Nacional
de Investigaciones Cient\'{\i}ficas y Tecnol\'{o}gicas (CONICYT).



\newpage
\section*{Figure captions}

\begin{description}
  \item [Figure 1:] \   \\
            Examples of fusions between two loops (a), between a
            loop and an open path (b) and between two open paths
            (c).
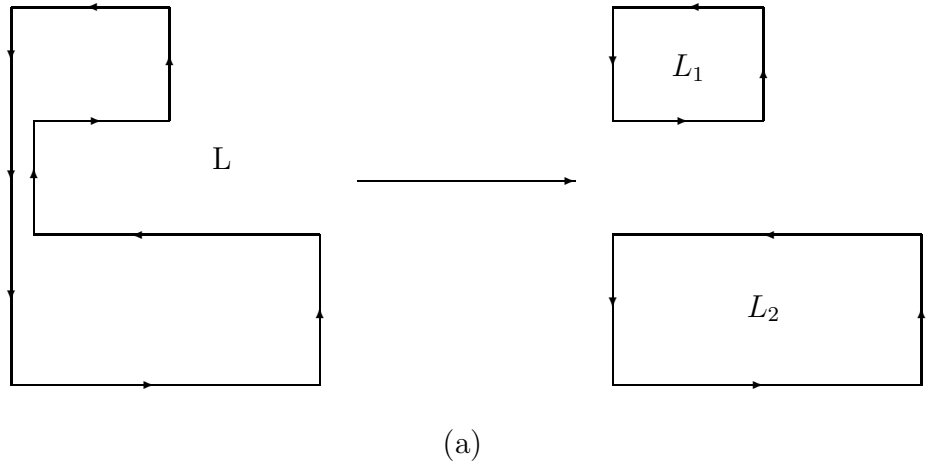
  \item [Figure 2:] \   \\
            (a) The fission term breaks the loop L and
                gives rise to the cluster $(L_{1}, L_{2})$. \\
            (b) In this case it breaks the open path P and gives
                rise to the cluster $(L, Q)$, where L is a loop and
                Q is an open path.
  \item [Figure 3:] \   \\
                The Mandelstam identities for two loops (a) and
                for one loop and one open path (b) with a common
                link.
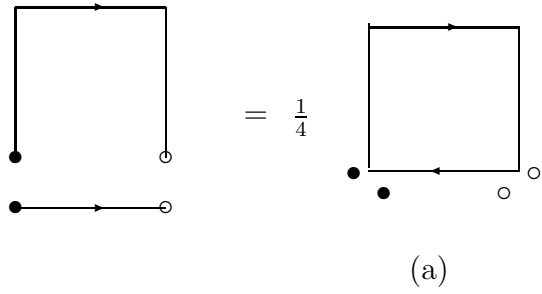
  \item [Figure 4:] \   \\
                The Mandelstam identities for two mesonic paths (a)
                and two barionic paths (b) with the same end
                points. The black circle represents one antiquark
                and the white circle one quark, two circles near
                represent two quarks placed on the same site.
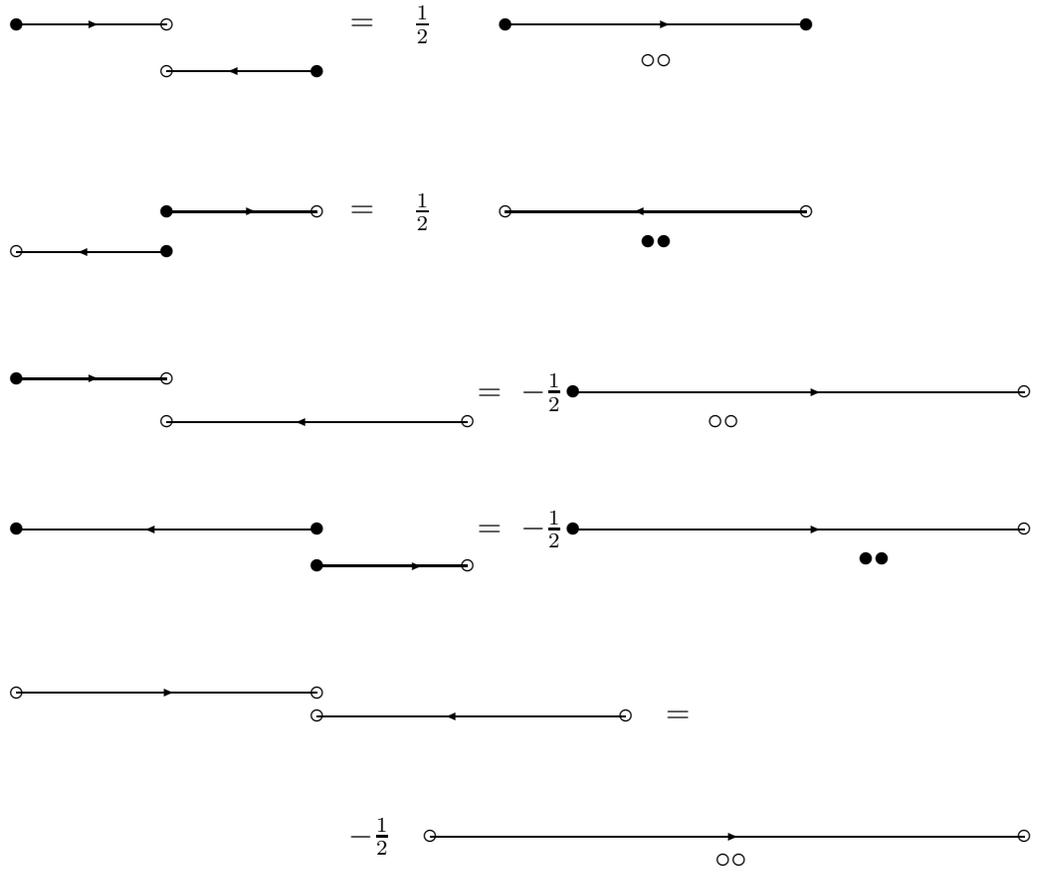
  \item [Figure 5:] \   \\
                The Mandelstam identities for two open paths with
                one equal end point.
  \item [Figure 6:] \   \\
                The basis of eleven clusters for $N_{max} = 2$.
\end{description}


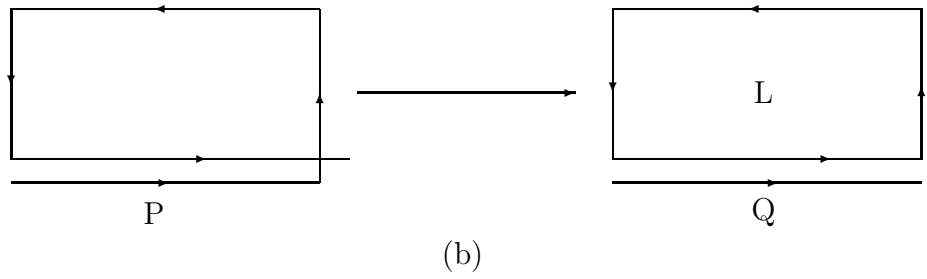
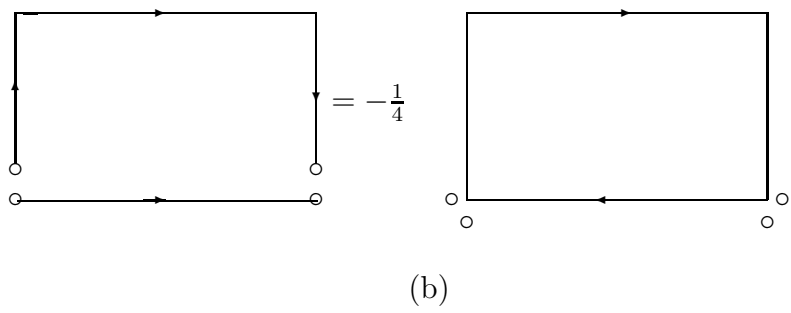
\unitlength=1.00mm
\special{em:linewidth 0.4pt}
\linethickness{0.4pt}
\begin{picture}(135.11,174.37)
\put(70.00,174.37){\makebox(0,0)[cc]{Fig. 1}}
\put(10.00,125.04){\framebox(41.00,20.00)[cc]{}}
\put(10.00,147.26){\framebox(21.00,16.89)[cc]{}}
\put(85.00,145.04){\rule{0.00\unitlength}{0.00\unitlength}}
\put(25.00,125.04){\vector(1,0){9.00}}
\put(51.00,131.26){\vector(0,1){4.89}}
\put(46.00,145.04){\vector(-1,0){6.00}}
\put(26.00,145.04){\vector(-1,0){6.00}}
\put(20.00,147.26){\vector(1,0){5.00}}
\put(31.00,151.26){\vector(0,1){4.89}}
\put(26.00,164.15){\vector(-1,0){6.00}}
\put(10.00,157.93){\vector(0,-1){4.00}}
\put(10.00,137.93){\vector(0,-1){4.00}}
\put(110.00,137.04){\makebox(0,0)[cc]{C}}
\put(29.00,135.26){\makebox(0,0)[cc]{B}}
\put(20.00,156.15){\makebox(0,0)[cc]{A}}
\put(70.00,115.26){\makebox(0,0)[cc]{(a)}}
\put(131.00,125.04){\line(-1,0){40.00}}
\put(91.00,125.04){\line(0,1){39.11}}
\put(91.00,164.15){\line(1,0){20.00}}
\put(111.00,164.15){\line(0,-1){18.22}}
\put(111.00,145.93){\line(1,0){20.00}}
\put(131.00,145.93){\line(0,-1){20.89}}
\put(105.00,125.04){\vector(1,0){6.00}}
\put(131.00,133.04){\vector(0,1){4.89}}
\put(126.00,145.93){\vector(-1,0){6.00}}
\put(111.00,149.93){\vector(0,1){5.33}}
\put(107.00,164.15){\vector(-1,0){5.00}}
\put(91.00,157.04){\vector(0,-1){4.89}}
\put(91.00,137.93){\vector(0,-1){4.00}}
\put(10.00,100.15){\line(1,0){41.00}}
\put(51.00,100.15){\line(0,-1){20.00}}
\put(51.00,80.15){\line(-1,0){41.00}}
\put(10.00,80.15){\line(0,1){20.00}}
\put(22.00,80.15){\vector(1,0){7.00}}
\put(10.00,76.15){\line(1,0){41.00}}
\put(26.00,76.15){\vector(1,0){9.00}}
\put(91.00,76.15){\line(1,0){40.00}}
\put(131.00,76.15){\line(0,1){24.00}}
\put(131.00,100.15){\line(-1,0){40.00}}
\put(91.00,100.15){\line(0,-1){20.00}}
\put(91.00,80.15){\line(1,0){44.00}}
\put(104.00,76.15){\vector(1,0){5.00}}
\put(109.00,80.15){\vector(1,0){7.00}}
\put(131.00,84.15){\vector(0,1){4.89}}
\put(118.00,100.15){\vector(-1,0){7.00}}
\put(91.00,93.04){\vector(0,-1){4.89}}
\put(70.00,69.93){\makebox(0,0)[cc]{(b)}}
\put(56.00,89.04){\vector(1,0){29.00}}
\put(56.00,145.93){\vector(1,0){30.00}}
\put(10.22,49.78){\line(0,-1){14.81}}
\put(10.22,34.96){\line(1,0){40.89}}
\put(51.11,34.96){\line(0,1){14.81}}
\put(51.11,14.81){\line(0,1){15.41}}
\put(51.11,30.22){\line(-1,0){40.89}}
\put(10.22,30.22){\line(0,-1){20.15}}
\put(51.11,10.07){\line(0,1){11.85}}
\put(51.11,40.30){\line(0,1){9.48}}
\put(51.11,46.81){\line(0,1){8.30}}
\put(10.22,55.11){\line(0,-1){8.30}}
\put(91.11,55.11){\line(0,-1){20.15}}
\put(91.11,34.96){\line(1,0){44.00}}
\put(135.11,34.96){\line(0,-1){24.89}}
\put(91.11,10.07){\line(0,1){20.15}}
\put(91.11,30.22){\line(1,0){40.00}}
\put(131.11,30.22){\line(0,1){24.89}}
\put(91.11,49.19){\vector(0,-1){4.15}}
\put(104.00,34.96){\vector(1,0){6.22}}
\put(135.11,30.22){\vector(0,-1){8.30}}
\put(91.11,17.78){\vector(0,1){2.96}}
\put(104.00,30.22){\vector(1,0){6.22}}
\put(131.11,40.30){\vector(0,1){3.56}}
\put(10.22,48.00){\vector(0,-1){4.15}}
\put(10.22,17.19){\vector(0,1){3.56}}
\put(24.00,30.22){\vector(1,0){8.00}}
\put(24.00,34.96){\vector(1,0){8.00}}
\put(51.11,40.89){\vector(0,1){5.33}}
\put(51.11,21.93){\vector(0,-1){5.93}}
\put(70.22,5.93){\makebox(0,0)[cc]{(c)}}
\put(56.00,33.04){\vector(1,0){29.00}}
\end{picture}


\unitlength=1.00mm
\special{em:linewidth 0.4pt}
\linethickness{0.4pt}
\begin{picture}(131.00,140.00)
\put(70.00,140.00){\makebox(0,0)[cc]{Fig. 2}}
\put(10.00,130.22){\line(1,0){21.00}}
\put(31.00,130.22){\line(0,-1){15.11}}
\put(31.00,115.11){\line(-1,0){18.00}}
\put(13.00,115.11){\line(0,-1){15.11}}
\put(13.00,100.00){\line(1,0){38.00}}
\put(51.00,100.00){\line(0,-1){20.00}}
\put(51.00,80.00){\line(-1,0){41.00}}
\put(10.00,80.00){\line(0,1){50.22}}
\put(90.00,130.22){\line(1,0){20.00}}
\put(110.00,130.22){\line(0,-1){15.11}}
\put(110.00,115.11){\line(-1,0){20.00}}
\put(90.00,115.11){\line(0,1){15.11}}
\put(90.00,100.00){\line(1,0){41.00}}
\put(131.00,100.00){\line(0,-1){20.00}}
\put(131.00,80.00){\line(-1,0){41.00}}
\put(90.00,80.00){\line(0,1){20.00}}
\put(110.00,90.22){\makebox(0,0)[cc]{$L_{2}$}}
\put(100.00,122.22){\makebox(0,0)[cc]{$L_{1}$}}
\put(38.00,110.22){\makebox(0,0)[cc]{L}}
\put(70.00,72.00){\makebox(0,0)[cc]{(a)}}
\put(106.00,80.00){\vector(1,0){4.00}}
\put(131.00,84.00){\vector(0,1){6.22}}
\put(114.00,100.00){\vector(-1,0){4.00}}
\put(90.00,95.11){\vector(0,-1){4.89}}
\put(95.00,115.11){\vector(1,0){5.00}}
\put(110.00,119.11){\vector(0,1){3.11}}
\put(106.00,130.22){\vector(-1,0){6.00}}
\put(90.00,127.11){\vector(0,-1){4.89}}
\put(31.00,120.00){\vector(0,1){4.00}}
\put(26.00,130.22){\vector(-1,0){6.00}}
\put(10.00,126.22){\vector(0,-1){3.11}}
\put(19.00,115.11){\vector(1,0){3.00}}
\put(10.00,111.11){\vector(0,-1){4.00}}
\put(13.00,107.11){\vector(0,1){1.78}}
\put(33.00,100.00){\vector(-1,0){7.00}}
\put(10.00,95.11){\vector(0,-1){4.00}}
\put(20.00,80.00){\vector(1,0){9.00}}
\put(51.00,84.89){\vector(0,1){5.33}}
\put(10.00,35.11){\line(0,1){20.00}}
\put(10.00,55.11){\line(1,0){41.00}}
\put(51.00,55.11){\line(0,-1){23.11}}
\put(51.00,32.00){\line(-1,0){41.00}}
\put(10.00,35.11){\line(1,0){45.00}}
\put(31.00,35.11){\vector(1,0){5.00}}
\put(24.00,32.00){\vector(1,0){7.00}}
\put(51.00,39.11){\vector(0,1){4.89}}
\put(38.00,55.11){\vector(-1,0){9.00}}
\put(10.00,50.22){\vector(0,-1){5.33}}
\put(90.00,55.11){\line(1,0){41.00}}
\put(131.00,55.11){\line(0,-1){20.00}}
\put(131.00,35.11){\line(-1,0){41.00}}
\put(90.00,35.11){\line(0,1){20.00}}
\put(90.00,32.00){\line(1,0){41.00}}
\put(108.00,32.00){\vector(1,0){4.00}}
\put(111.00,35.11){\vector(1,0){8.00}}
\put(131.00,40.89){\vector(0,1){4.00}}
\put(116.00,55.11){\vector(-1,0){8.00}}
\put(90.00,50.22){\vector(0,-1){6.22}}
\put(56.00,44.00){\vector(1,0){29.00}}
\put(56.00,107.11){\vector(1,0){29.00}}
\put(70.00,22.22){\makebox(0,0)[cc]{(b)}}
\put(29.00,28.00){\makebox(0,0)[cc]{P}}
\put(110.00,28.00){\makebox(0,0)[cc]{Q}}
\put(110.00,44.00){\makebox(0,0)[cc]{L}}
\end{picture}


\unitlength=1.00mm
\special{em:linewidth 0.4pt}
\linethickness{0.4pt}
\begin{picture}(145.00,140.00)
\put(70.00,140.00){\makebox(0,0)[cc]{Fig. 3}}
\put(10.00,130.22){\line(1,0){15.00}}
\put(25.00,130.22){\line(0,-1){35.11}}
\put(25.00,95.11){\line(1,0){20.00}}
\put(45.00,95.11){\line(0,1){15.11}}
\put(45.00,110.22){\line(-1,0){17.00}}
\put(28.00,110.22){\line(0,-1){12.00}}
\put(28.00,98.22){\line(-1,0){5.00}}
\put(23.00,98.22){\line(0,-1){3.11}}
\put(23.00,95.11){\line(-1,0){13.00}}
\put(10.00,95.11){\line(0,1){35.11}}
\put(55.00,112.00){\makebox(0,0)[cc]{  =    }}
\put(65.00,130.22){\line(1,0){16.00}}
\put(81.00,130.22){\line(0,-1){35.11}}
\put(81.00,95.11){\line(-1,0){16.00}}
\put(65.00,95.11){\line(0,1){35.11}}
\put(84.00,110.22){\line(1,0){17.00}}
\put(101.00,110.22){\line(0,-1){15.11}}
\put(101.00,95.11){\line(-1,0){17.00}}
\put(84.00,95.11){\line(0,1){15.11}}
\put(105.00,112.00){\makebox(0,0)[cc]{$-$ }}
\put(111.00,130.22){\line(1,0){14.00}}
\put(125.00,130.22){\line(0,-1){20.00}}
\put(125.00,110.22){\line(1,0){20.00}}
\put(145.00,110.22){\line(0,-1){15.11}}
\put(145.00,95.11){\line(-1,0){34.00}}
\put(111.00,95.11){\line(0,1){35.11}}
\put(78.00,102.22){\makebox(0,0)[cc]{$\ell$}}
\put(22.00,102.22){\makebox(0,0)[cc]{$\ell$}}
\put(25.00,90.22){\makebox(0,0)[cc]{$(A \ell B \ell)$}}
\put(80.00,83.11){\makebox(0,0)[cc]{(a)}}
\put(48.00,40.00){\line(-1,0){38.00}}
\put(10.00,40.00){\line(0,1){20.00}}
\put(10.00,60.00){\line(1,0){35.00}}
\put(45.00,60.00){\line(0,0){0.00}}
\put(45.00,60.00){\line(0,-1){24.00}}
\put(45.00,36.00){\line(-1,0){35.00}}
\put(21.00,36.00){\vector(1,0){6.00}}
\put(25.00,40.00){\vector(1,0){7.00}}
\put(45.00,44.89){\vector(0,1){4.00}}
\put(32.00,60.00){\vector(-1,0){7.00}}
\put(10.00,52.89){\vector(0,-1){4.00}}
\put(55.00,50.22){\makebox(0,0)[cc]{  =    }}
\put(65.00,60.00){\line(1,0){36.00}}
\put(101.00,60.00){\line(0,0){0.00}}
\put(101.00,60.00){\line(0,-1){20.00}}
\put(101.00,40.00){\line(-1,0){36.00}}
\put(65.00,40.00){\line(0,1){20.00}}
\put(87.00,60.00){\vector(-1,0){8.00}}
\put(65.00,54.22){\vector(0,-1){5.33}}
\put(75.00,40.00){\vector(1,0){7.00}}
\put(101.00,46.22){\vector(0,1){4.00}}
\put(101.00,36.00){\line(-1,0){36.00}}
\put(76.00,36.00){\vector(1,0){5.00}}
\put(105.00,50.22){\makebox(0,0)[cc]{$-$ }}
\put(111.00,40.00){\line(0,1){20.00}}
\put(111.00,60.00){\line(1,0){34.00}}
\put(145.00,60.00){\line(0,-1){20.00}}
\put(145.00,52.89){\vector(0,-1){4.89}}
\put(124.00,60.00){\vector(1,0){7.00}}
\put(111.00,47.11){\vector(0,1){4.89}}
\put(118.00,95.11){\vector(1,0){6.00}}
\put(124.00,95.11){\vector(1,0){13.00}}
\put(145.00,98.22){\vector(0,1){4.89}}
\put(140.00,110.22){\vector(-1,0){6.00}}
\put(125.00,115.11){\vector(0,1){5.78}}
\put(123.00,130.22){\vector(-1,0){5.00}}
\put(111.00,126.22){\vector(0,-1){5.33}}
\put(111.00,107.11){\vector(0,-1){4.00}}
\put(101.00,104.89){\vector(0,-1){2.67}}
\put(96.00,95.11){\vector(-1,0){4.00}}
\put(84.00,99.11){\vector(0,1){4.00}}
\put(81.00,99.11){\vector(0,1){5.78}}
\put(89.00,110.22){\vector(1,0){5.00}}
\put(81.00,116.00){\vector(0,1){4.89}}
\put(77.00,130.22){\vector(-1,0){6.00}}
\put(65.00,112.89){\vector(0,0){0.00}}
\put(70.00,95.11){\vector(1,0){5.00}}
\put(39.00,95.11){\vector(-1,0){5.00}}
\put(16.00,95.11){\vector(1,0){4.00}}
\put(10.00,118.22){\vector(0,-1){8.00}}
\put(20.00,130.22){\vector(-1,0){4.00}}
\put(25.00,104.00){\vector(0,1){4.00}}
\put(28.00,100.89){\vector(0,1){3.11}}
\put(32.00,110.22){\vector(1,0){5.00}}
\put(45.00,106.22){\vector(0,-1){4.00}}
\put(80.00,24.00){\makebox(0,0)[cc]{(b)}}
\end{picture}


\unitlength=1.00mm
\special{em:linewidth 0.4pt}
\linethickness{0.4pt}
\begin{picture}(112.00,140.00)
\put(65.00,140.00){\makebox(0,0)[cc]{Fig.  4}}
\put(10.00,110.22){\line(0,1){20.00}}
\put(10.00,130.22){\line(1,0){20.00}}
\put(30.00,130.22){\line(0,-1){20.00}}
\put(30.00,110.22){\makebox(0,0)[cc]{$\circ$}}
\put(10.00,110.22){\makebox(0,0)[cc]{$\bullet$}}
\put(10.00,103.56){\line(1,0){20.00}}
\put(30.00,103.56){\makebox(0,0)[cc]{$\circ$}}
\put(10.00,103.56){\makebox(0,0)[cc]{$\bullet$}}
\put(42.00,115.56){\makebox(0,0)[cc]{$ = $}}
\put(57.00,127.56){\line(1,0){20.00}}
\put(77.00,127.56){\line(0,-1){19.11}}
\put(77.00,108.44){\line(-1,0){20.00}}
\put(48.00,115.56){\makebox(0,0)[cc]{$\frac{1}{4}$}}
\put(55.00,108.00){\makebox(0,0)[cc]{$\bullet$}}
\put(59.00,105.33){\makebox(0,0)[cc]{$\bullet$}}
\put(75.00,105.33){\makebox(0,0)[cc]{$\circ$}}
\put(79.00,108.00){\makebox(0,0)[cc]{$\circ$}}
\put(18.00,103.56){\vector(1,0){4.00}}
\put(18.00,130.22){\vector(1,0){4.00}}
\put(64.00,127.56){\vector(1,0){5.00}}
\put(69.00,108.44){\vector(-1,0){4.00}}
\put(50.22,50.07){\line(-1,0){39.11}}
\put(11.11,74.96){\line(1,0){1.78}}
\put(10.22,50.07){\line(1,0){5.78}}
\put(27.00,50.22){\vector(1,0){3.00}}
\put(10.00,55.11){\line(0,1){20.00}}
\put(26.00,75.11){\vector(1,0){4.00}}
\put(10.00,60.44){\vector(0,1){5.78}}
\put(50.00,75.11){\line(0,-1){20.00}}
\put(50.00,68.44){\vector(0,-1){5.33}}
\put(10.00,73.33){\line(0,1){1.78}}
\put(10.00,75.11){\line(1,0){40.00}}
\put(50.00,75.11){\line(0,-1){4.00}}
\put(50.00,54.22){\makebox(0,0)[cc]{$\circ$}}
\put(10.00,54.22){\makebox(0,0)[cc]{$\circ$}}
\put(10.00,50.22){\makebox(0,0)[cc]{$\circ$}}
\put(50.00,50.22){\makebox(0,0)[cc]{$\circ$}}
\put(57.00,63.11){\makebox(0,0)[cc]{$ = -\frac{1}{4}$}}
\put(70.00,50.22){\line(1,0){40.00}}
\put(110.00,50.22){\line(0,1){24.89}}
\put(110.00,75.11){\line(-1,0){40.00}}
\put(70.00,75.11){\line(0,-1){24.89}}
\put(68.00,50.22){\makebox(0,0)[cc]{$\circ$}}
\put(70.00,47.11){\makebox(0,0)[cc]{$\circ$}}
\put(110.00,47.11){\makebox(0,0)[cc]{$\circ$}}
\put(112.00,50.22){\makebox(0,0)[cc]{$\circ$}}
\put(91.00,50.22){\vector(-1,0){4.00}}
\put(87.00,75.11){\vector(1,0){5.00}}
\put(57.00,128.00){\line(0,-1){19.11}}
\put(65.00,95.11){\makebox(0,0)[cc]{(a)}}
\put(65.00,38.22){\makebox(0,0)[cc]{(b)}}
\end{picture}


\unitlength=1.00mm
\special{em:linewidth 0.4pt}
\linethickness{0.4pt}
\begin{picture}(144.00,140.00)
\put(65.00,140.00){\makebox(0,0)[cc]{Fig.  5}}
\put(10.00,127.11){\line(1,0){20.00}}
\put(30.00,120.89){\line(1,0){20.00}}
\put(50.00,120.89){\makebox(0,0)[cc]{$\bullet$}}
\put(30.00,120.89){\makebox(0,0)[cc]{$\circ$}}
\put(30.00,127.11){\makebox(0,0)[cc]{$\circ$}}
\put(10.00,127.11){\makebox(0,0)[cc]{$\bullet$}}
\put(17.00,127.11){\vector(1,0){4.00}}
\put(42.00,120.89){\vector(-1,0){4.00}}
\put(56.00,127.11){\makebox(0,0)[cc]{$ = $}}
\put(64.00,127.11){\makebox(0,0)[cc]{$\frac{1}{2}$}}
\put(75.00,127.11){\line(1,0){40.00}}
\put(115.00,127.11){\makebox(0,0)[cc]{$\bullet$}}
\put(95.00,122.22){\makebox(0,0)[cc]{$\circ \circ$}}
\put(90.00,127.11){\vector(1,0){7.00}}
\put(75.00,127.11){\makebox(0,0)[cc]{$\bullet$}}
\put(30.00,102.22){\line(1,0){20.00}}
\put(50.00,102.22){\makebox(0,0)[cc]{$\circ$}}
\put(30.00,102.22){\makebox(0,0)[cc]{$\bullet$}}
\put(30.00,96.89){\line(-1,0){20.00}}
\put(10.00,96.89){\makebox(0,0)[cc]{$\circ$}}
\put(30.00,96.89){\makebox(0,0)[cc]{$\bullet$}}
\put(22.00,96.89){\vector(-1,0){4.00}}
\put(36.00,102.22){\vector(1,0){6.00}}
\put(56.00,102.22){\makebox(0,0)[cc]{$ = $}}
\put(64.00,102.22){\makebox(0,0)[cc]{$\frac{1}{2}$}}
\put(75.00,102.22){\line(1,0){40.00}}
\put(75.00,102.22){\makebox(0,0)[cc]{$\circ$}}
\put(115.00,102.22){\makebox(0,0)[cc]{$\circ$}}
\put(95.00,98.22){\makebox(0,0)[cc]{$\bullet \bullet$}}
\put(97.00,102.22){\vector(-1,0){5.00}}
\put(10.00,80.00){\line(1,0){20.00}}
\put(30.00,74.22){\line(1,0){40.00}}
\put(77.00,78.22){\makebox(0,0)[cc]{$ = \ - \frac{1}{2}$}}
\put(84.00,78.22){\line(1,0){60.00}}
\put(144.00,78.22){\makebox(0,0)[cc]{$\circ$}}
\put(104.00,74.22){\makebox(0,0)[cc]{$\circ \circ$}}
\put(112.00,78.22){\vector(1,0){5.00}}
\put(84.00,78.22){\makebox(0,0)[cc]{$\bullet$}}
\put(70.00,74.22){\makebox(0,0)[cc]{$\circ$}}
\put(51.00,74.22){\vector(-1,0){4.00}}
\put(30.00,74.22){\makebox(0,0)[cc]{$\circ$}}
\put(30.00,80.00){\makebox(0,0)[cc]{$\circ$}}
\put(17.00,80.00){\vector(1,0){4.00}}
\put(10.00,80.00){\makebox(0,0)[cc]{$\bullet$}}
\put(50.00,55.11){\line(1,0){20.00}}
\put(70.00,55.11){\makebox(0,0)[cc]{$\circ$}}
\put(59.00,55.11){\vector(1,0){5.00}}
\put(50.00,55.11){\makebox(0,0)[cc]{$\bullet$}}
\put(50.00,60.00){\makebox(0,0)[cc]{$\bullet$}}
\put(10.00,60.00){\makebox(0,0)[cc]{$\bullet$}}
\put(77.00,60.00){\makebox(0,0)[cc]{$ = \ - \frac{1}{2}$}}
\put(10.00,60.00){\line(1,0){40.00}}
\put(31.00,60.00){\vector(-1,0){4.00}}
\put(84.00,60.00){\line(1,0){60.00}}
\put(144.00,60.00){\makebox(0,0)[cc]{$\circ$}}
\put(112.00,60.00){\vector(1,0){5.00}}
\put(124.00,56.00){\makebox(0,0)[cc]{$\bullet \bullet$}}
\put(84.00,60.00){\makebox(0,0)[cc]{$\bullet$}}
\put(10.00,38.22){\line(1,0){40.00}}
\put(50.00,35.11){\line(1,0){41.00}}
\put(91.00,35.11){\makebox(0,0)[cc]{$\circ$}}
\put(50.00,35.11){\makebox(0,0)[cc]{$\circ$}}
\put(50.00,38.22){\makebox(0,0)[cc]{$\circ$}}
\put(10.00,38.22){\makebox(0,0)[cc]{$\circ$}}
\put(25.00,38.22){\vector(1,0){6.00}}
\put(74.00,35.11){\vector(-1,0){7.00}}
\put(98.00,35.11){\makebox(0,0)[cc]{$ = $}}
\put(144.00,19.11){\line(-1,0){39.00}}
\put(105.00,19.11){\line(-1,0){40.00}}
\put(57.00,19.11){\makebox(0,0)[cc]{$- \frac{1}{2}$}}
\put(65.00,19.11){\makebox(0,0)[cc]{$\circ$}}
\put(144.00,19.11){\makebox(0,0)[cc]{$\circ$}}
\put(102.00,19.11){\vector(1,0){4.00}}
\put(105.00,16.00){\makebox(0,0)[cc]{$\circ \circ$}}
\end{picture}


\unitlength=1.00mm
\special{em:linewidth 0.4pt}
\linethickness{0.4pt}
\begin{picture}(130.00,164.89)
\put(65.00,164.89){\makebox(0,0)[cc]{Fig.  6}}
\put(5.00,155.11){\makebox(0,0)[cc]{{\bf Order 1:}}}
\put(10.00,130.22){\line(1,0){20.00}}
\put(30.00,130.22){\makebox(0,0)[cc]{$\circ$}}
\put(10.00,130.22){\makebox(0,0)[cc]{$\bullet$}}
\put(17.00,130.22){\vector(1,0){5.00}}
\put(50.00,130.22){\line(1,0){20.00}}
\put(70.00,130.22){\line(0,1){20.00}}
\put(70.00,150.22){\line(-1,0){20.00}}
\put(50.00,150.22){\line(0,-1){20.00}}
\put(60.00,124.89){\makebox(0,0)[cc]{2}}
\put(20.00,124.89){\makebox(0,0)[cc]{1}}
\put(5.00,115.11){\makebox(0,0)[cc]{{\bf Order 2:}}}
\put(10.00,92.89){\makebox(0,0)[cc]{$\bullet$}}
\put(10.00,87.11){\makebox(0,0)[cc]{$\bullet$}}
\put(30.00,92.89){\makebox(0,0)[cc]{$\circ$}}
\put(30.00,87.11){\makebox(0,0)[cc]{$\circ$}}
\put(50.00,90.22){\line(1,0){20.00}}
\put(70.00,90.22){\line(0,1){20.00}}
\put(70.00,110.22){\line(-1,0){20.00}}
\put(50.00,110.22){\line(0,-1){20.00}}
\put(50.00,87.11){\line(1,0){20.00}}
\put(58.00,87.11){\vector(1,0){4.00}}
\put(60.00,82.22){\makebox(0,0)[cc]{4}}
\put(70.00,87.11){\makebox(0,0)[cc]{$\circ$}}
\put(50.00,87.11){\makebox(0,0)[cc]{$\bullet$}}
\put(20.00,82.22){\makebox(0,0)[cc]{3}}
\put(90.00,110.22){\line(1,0){40.00}}
\put(130.00,110.22){\makebox(0,0)[cc]{$\bullet$}}
\put(90.00,110.22){\makebox(0,0)[cc]{$\bullet$}}
\put(106.00,110.22){\vector(1,0){4.00}}
\put(110.00,107.11){\makebox(0,0)[cc]{$\circ \circ$}}
\put(110.00,102.22){\makebox(0,0)[cc]{5}}
\put(130.00,90.22){\line(-1,0){40.00}}
\put(90.00,90.22){\makebox(0,0)[cc]{$\circ$}}
\put(105.00,90.22){\vector(1,0){5.00}}
\put(110.00,87.11){\makebox(0,0)[cc]{$\bullet \bullet$}}
\put(110.00,82.22){\makebox(0,0)[cc]{6}}
\put(10.00,50.22){\line(1,0){20.00}}
\put(30.00,50.22){\line(0,1){20.00}}
\put(30.00,70.22){\makebox(0,0)[cc]{$\bullet$}}
\put(30.00,56.89){\vector(0,1){3.11}}
\put(32.00,50.22){\makebox(0,0)[cc]{$\circ$}}
\put(30.00,47.11){\makebox(0,0)[cc]{$\circ$}}
\put(17.00,50.22){\vector(1,0){3.00}}
\put(10.00,50.22){\makebox(0,0)[cc]{$\bullet$}}
\put(20.00,44.00){\makebox(0,0)[cc]{7}}
\put(70.00,50.22){\line(-1,0){20.00}}
\put(50.00,50.22){\line(0,1){20.00}}
\put(50.00,70.22){\makebox(0,0)[cc]{$\circ$}}
\put(50.00,56.89){\vector(0,1){3.11}}
\put(48.00,50.22){\makebox(0,0)[cc]{$\bullet$}}
\put(50.00,47.11){\makebox(0,0)[cc]{$\bullet$}}
\put(63.00,50.22){\vector(-1,0){3.00}}
\put(60.00,44.89){\makebox(0,0)[cc]{8}}
\put(130.00,90.22){\makebox(0,0)[cc]{$\circ$}}
\put(70.00,50.22){\makebox(0,0)[cc]{$\circ$}}
\put(30.00,90.22){\line(-1,0){2.00}}
\put(26.00,90.22){\line(-1,0){2.00}}
\put(22.00,90.22){\line(-1,0){2.00}}
\put(18.00,90.22){\line(-1,0){2.00}}
\put(14.00,90.22){\line(-1,0){2.00}}
\put(10.00,90.22){\line(0,1){0.00}}
\put(11.00,90.22){\line(-1,0){1.00}}
\put(90.00,50.00){\line(1,0){20.00}}
\put(110.00,50.00){\line(0,1){20.00}}
\put(110.00,70.00){\line(-1,0){19.17}}
\put(92.92,52.22){\line(1,0){15.00}}
\put(107.92,52.22){\line(0,1){15.56}}
\put(107.92,67.78){\line(-1,0){15.00}}
\put(92.92,67.78){\line(0,-1){15.56}}
\put(100.00,45.00){\makebox(0,0)[cc]{9}}
\put(10.00,30.00){\line(1,0){20.00}}
\put(30.00,30.00){\line(0,-1){20.00}}
\put(30.00,10.00){\line(-1,0){20.00}}
\put(10.00,10.00){\line(0,1){20.00}}
\put(32.08,30.00){\line(1,0){20.00}}
\put(52.08,30.00){\line(0,-1){20.00}}
\put(52.08,10.00){\line(-1,0){20.00}}
\put(32.08,10.00){\line(0,1){20.00}}
\put(30.83,6.11){\makebox(0,0)[cc]{10}}
\put(72.92,30.00){\line(1,0){19.17}}
\put(92.08,30.00){\line(0,-1){20.00}}
\put(92.08,10.00){\line(-1,0){21.25}}
\put(70.83,10.00){\line(0,1){20.00}}
\put(70.83,30.00){\line(1,0){21.25}}
\put(94.17,30.00){\line(0,-1){20.00}}
\put(94.17,10.00){\line(5,3){17.92}}
\put(112.08,20.56){\line(0,1){19.44}}
\put(92.92,6.11){\makebox(0,0)[cc]{11}}
\put(90.00,70.22){\line(0,-1){20.00}}
\put(90.00,70.22){\line(1,0){7.00}}
\put(112.00,40.89){\line(-5,-3){18.00}}
\end{picture}

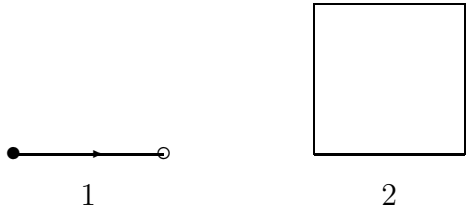
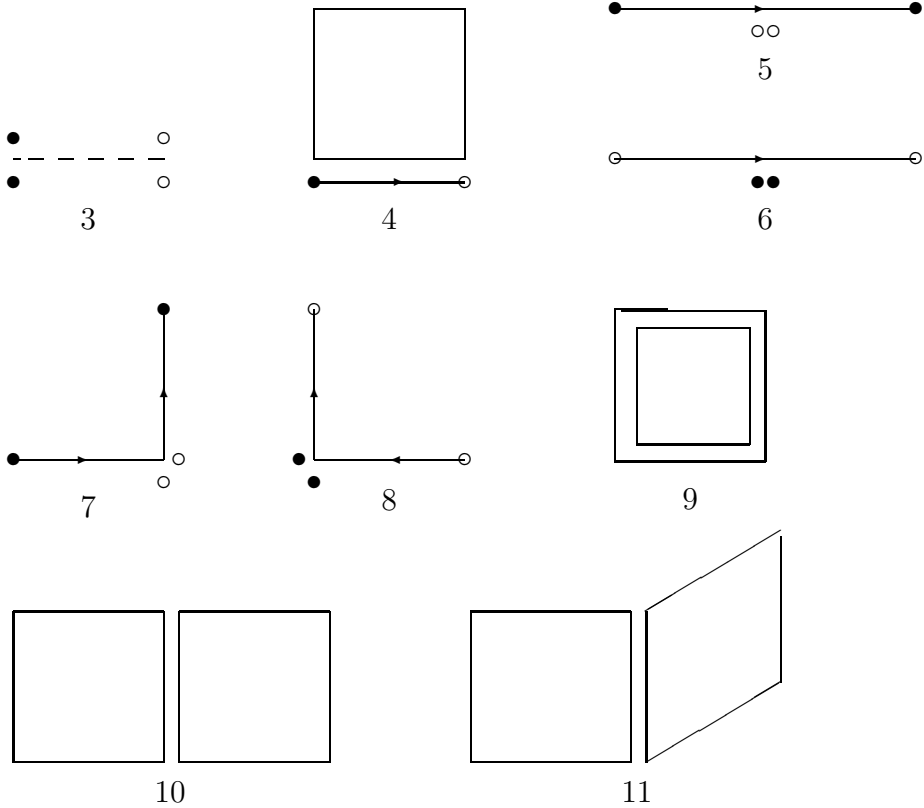
\end{document}